\documentclass[twocolumn]{aastex62}

\usepackage{mathtools}
\usepackage{xfrac}
\usepackage{multirow}
\usepackage{color}
\usepackage{xcolor}
\usepackage{xspace}

\newcommand{\RA}[4]{{#1}$^\text{h}${#2}$^\text{m}${#3}$^\text{s}${#4}}
\newcommand{\dec}[4]{{#1}\textdegree {#2}'{#3}''{#4}}
\newcommand{\OH}{$12 + \log(\text{O}/\text{H})$\xspace}
\newcommand{\NH}{$12 + \log(\text{N}/\text{H})$\xspace}
\newcommand{\NeH}{$12 + \log(\text{Ne}/\text{H})$\xspace}
\newcommand{\NO}{$\log(\text{N}/\text{O})$\xspace}
\newcommand{\NeO}{$\log(\text{Ne}/\text{O})$\xspace}

\shorttitle{O$^+$ approximation}
\shortauthors{Douglass, Vogeley, and Cen}

\begin{document}

%%%%%%%%%%%%%%%%%%%%%%%%%%%%%%%%%%%%%%%%%%%%%%%%%%%%%%%%%%%%%%%%%%%%%%%%%%%%%%%%
%
%    TITLE
%
%%%%%%%%%%%%%%%%%%%%%%%%%%%%%%%%%%%%%%%%%%%%%%%%%%%%%%%%%%%%%%%%%%%%%%%%%%%%%%%%
\title{Influence of the Void Environment on Chemical Abundances in Dwarf Galaxies and Implications for Connecting Star Formation and Halo Mass}

\correspondingauthor{Kelly A. Douglass}
\email{kelly.a.douglass@drexel.edu}

\author[0000-0002-9540-546X]{Kelly A. Douglass}
\altaffiliation{Department of Physics \& Astronomy, University of Rochester\\
500 Wilson Blvd., Rochester, NY  14611}
\affiliation{Department of Physics, Drexel University, 3141 Chestnut Street, Philadelphia, PA  19104}

\author[0000-0001-7416-9800]{Michael S. Vogeley}
\affiliation{Department of Physics, Drexel University, 3141 Chestnut Street, Philadelphia, PA  19104}

\author{Renyue Cen}
\affiliation{Department of Astrophysics, Princeton University, Peyton Hall, Princeton, NJ  05844}

%%%%%%%%%%%%%%%%%%%%%%%%%%%%%%%%%%%%%%%%%%%%%%%%%%%%%%%%%%%%%%%%%%%%%%%%%%%%%%%%
%
%    ABSTRACT
%
%%%%%%%%%%%%%%%%%%%%%%%%%%%%%%%%%%%%%%%%%%%%%%%%%%%%%%%%%%%%%%%%%%%%%%%%%%%%%%%%
\begin{abstract}
We study how the void environment affects the chemical evolution of galaxies in 
the universe by comparing the oxygen and nitrogen abundances of dwarf galaxies 
in voids with dwarf galaxies in denser regions.  Using spectroscopic 
observations from the Sloan Digital Sky Survey Data Release 7, we estimate the 
oxygen, nitrogen, and neon abundances of 889 void dwarf galaxies and 672 dwarf 
galaxies in denser regions.  We use the Direct $T_e$ method for calculating the 
gas-phase chemical abundances in the dwarf galaxies because it is best suited 
for low-metallicity, low-mass (dwarf) galaxies.  A substitute for the 
[\ion{O}{2}] $\lambda 3727$ doublet is developed, permitting oxygen abundance 
estimates of SDSS dwarf galaxies at all redshifts with the Direct $T_e$ method.  
We find that void dwarf galaxies have about the same oxygen abundance and Ne/O 
ratio as dwarf galaxies in denser environments.  However, we find that void 
dwarf galaxies have slightly higher neon ($\sim$10\%) abundances than dwarf 
galaxies in denser environments.  The opposite trend is seen in both the 
nitrogen abundance and N/O ratio: void dwarf galaxies have slightly lower 
nitrogen abundances ($\sim$5\%) and lower N/O ratios ($\sim$7\%) than dwarf 
galaxies in denser regions.  Therefore, we conclude that the void environment 
has a slight influence on dwarf galaxy chemical evolution.  Our mass--N/O 
relationship shows that the secondary production of nitrogen commences at a 
lower stellar mass in void dwarf star-forming galaxies than in dwarf 
star-forming galaxies in denser environments.  We also find that star-forming 
void dwarf galaxies have higher \ion{H}{1} masses than the star-forming dwarf 
galaxies in denser regions.  Our star-forming dwarf galaxy sample demonstrates a 
strong anti-correlation between the sSFR and N/O ratio, providing evidence that 
oxygen is produced in higher-mass stars than those which synthesize nitrogen.  
The lower N/O ratios and smaller stellar mass for secondary nitrogen production 
seen in void dwarf galaxies may indicate both delayed star formation as 
predicted by $\Lambda$CDM cosmology and a dependence of cosmic downsizing on the 
large-scale environment.  A shift toward slightly higher oxygen abundances and 
higher \ion{H}{1} masses in void dwarf galaxies could be evidence of larger 
ratios of dark matter halo mass to stellar mass in voids compared with denser 
regions.
\end{abstract}

\keywords{galaxies: abundances --- galaxies: dwarf --- galaxies: evolution}

%%%%%%%%%%%%%%%%%%%%%%%%%%%%%%%%%%%%%%%%%%%%%%%%%%%%%%%%%%%%%%%%%%%%%%%%%%%%%%%%
%
%    INTRODUCTION
%
%%%%%%%%%%%%%%%%%%%%%%%%%%%%%%%%%%%%%%%%%%%%%%%%%%%%%%%%%%%%%%%%%%%%%%%%%%%%%%%%
\section{Introduction}

Galactic redshift surveys have revealed that the large-scale distribution of 
galaxies is similar to a three-dimensional cosmic web \citep{Bond96}, with thin 
filaments of galaxies connecting galaxy clusters separated by voids (large, 
underdense areas that fill more than 60\% of space).  The voids first identified 
in early surveys \citep[e.g.,][]{Gregory78,Kirshner81,deLapparent86} have proven 
to be a universal feature of large-scale structure.  Analyses of the Sloan 
Digital Sky Survey \citep{Abazajian09,Ahn12} have produced catalogs of $10^3$ 
voids \citep{Pan12,Sutter14}.  Cosmic voids are an essential component for 
understanding the role of a galaxy's environment on its formation and evolution 
\citep[see][for a review]{vandeWeygaert11}.  

Extensive studies have been performed to understand the role of the environment 
in galaxy formation.  A strong relationship was found between a galaxy's 
morphology and the local density \citep{Dressler80,Postman84}, where the 
fraction of early-type galaxies increases with density.  A galaxy's luminosity 
was found to also contribute to this morphology--luminosity--density relation 
\citep{Park07}.  While much of this early work focused on trends of galaxy 
properties in the densest regions of space, evidence was found that the same 
trends persist into the voids, where galaxies are found to be bluer 
\citep{Grogin99,Rojas04,Patiri06,vonBendaBeckmann08,Hoyle12}, to be of a later 
morphological type \citep{Grogin00,Rojas04,Park07}, and to have higher specific 
star formation rates \citep[sSFR;][]{Rojas05,vonBendaBeckmann08,Moorman15,
Beygu16}.  These trends are attributed to the availability of cool gas to feed 
star formation in the void regions.  \cite{Hoyle05} and \cite{Moorman15} showed 
that there is a shift toward fainter objects in the void galaxy luminosity 
function.  This shift is consistent with the predicted shift of the dark matter 
halo mass function \citep{Goldberg05}.  Investigations into the \ion{H}{1} 
properties of void galaxies have also been performed \citep{Kreckel12,
Moorman14}, where void galaxies tend to have lower \ion{H}{1} masses than 
galaxies in denser environments.  All these observations are consistent with 
predictions from the $\Lambda$CDM cosmology that void galaxies have lower masses 
and are retarded in their star formation when compared to those in denser 
environments \citep[e.g.,][]{Gottlober03,Goldberg05,Cen11}.

Given that the sSFR and evolutionary history are different for galaxies in 
voids, it follows that their chemical evolution might also be influenced by the 
environment.  The metallicity of a galaxy (a measure of the integrated star 
formation history) is an estimate of the percentage of the galaxy's gas that has 
been processed in stars \citep{Guseva09}.  We would expect void galaxies to have 
lower metallicities than those in denser regions if they have only recently 
commenced forming stars or have recently accreted unprocessed gas.  Observations 
by \cite{Cooper08,Deng11,Filho15,Pustilnik06,Pustilnik11a,Pustilnik11b,
Pustilnik13,Pustilnik14}, and \cite{Pilyugin17} support the hypothesis that void 
galaxies have lower metallicities than galaxies in denser regions.  However, 
\cite{Kreckel15} and \cite{Douglass17a} find no influence from the large-scale 
environment on the metallicity, and \cite{Douglass17b} find that void dwarf 
galaxies have higher metallicities than dwarf galaxies in denser regions.  It is 
obvious that a study of a statistically significant large sample of galaxies is 
required to understand how the large-scale environment influences the chemical 
evolution of galaxies.

Environmental effects should be the most obvious on dwarf galaxies, since they 
possess small gravitational potential wells.  As a result, they are more 
sensitive to astrophysical effects such as cosmological reionization, internal 
feedback from supernova and photoheating from star formation, external effects 
from tidal interactions and ram pressure stripping, small-scale details of dark 
matter halo assembly, and properties of dark matter.  The main galaxy sample of 
SDSS DR7 covers a large enough volume to identify more than 1000 voids 
\citep{Pan12}, along with a statistically significant sample of dwarf galaxies 
($M_r > -17$) in voids.  SDSS also provides spectroscopy that permits 
metallicity estimates of this large sample of dwarf galaxies.

Numerous methods to estimate the metallicity of an object have been developed 
over the years, as a result of the availability of various spectral features.  
All methods except the direct $T_e$ method are calibrated on galaxies with 
various characteristics, making it unwise to apply them to galaxies outside the 
groups from which these calibrations were derived.  The commonly used methods 
are not calibrated for low-mass galaxies, so we carefully chose to use the 
Direct $T_e$ method because we are focusing on only dwarf galaxies.  A detailed 
explanation of this and other method classes can be found in \cite{Douglass17a}.  
With this estimator, \cite{Douglass17a, Douglass17b} have looked at the 
gas-phase chemical abundances of 135 star-forming dwarf galaxies.  They found 
that the large-scale environment has very little influence on the oxygen and 
nitrogen abundances in star-forming dwarf galaxies.  However, their sample does 
indicate that star-forming void galaxies have lower N/O ratios than star-forming 
dwarf galaxies in denser regions.  This is attributed to delayed star formation 
in void galaxies, along with a possible environmental influence on cosmic 
downsizing.  They also argue that the very slight shift toward higher oxygen 
abundances in star-forming void dwarf galaxies could be due to a larger ratio of 
dark matter halo mass to stellar mass in void dwarf galaxies.

We look to expand on the work by \cite{Douglass17a,Douglass17b} by substantially 
increasing their sample size of dwarf galaxies.  The main limiting factor in 
their sample was due to the required detection of the [\ion{O}{2}] $\lambda$3727 
doublet, which is needed to estimate the amount of singly ionized oxygen 
present.  We present a new approach to the O$^+$ abundance estimation, which 
removes the need for this emission line.  This calculation is used in 
conjunction with the Direct $T_e$ method to estimate the total gas-phase oxygen 
abundance in galaxies.  We also examine the relationship between the chemical 
abundance and \ion{H}{1} mass of star-forming dwarf galaxies.  For reasons 
previously described, this work only examines the chemical evolution of 
star-forming dwarf galaxies.  As a result of this sample, the known influences 
of the environment on the morphology and luminosity have already been taken into 
account, since we concentrate only on star-forming dwarf galaxies.

%%%%%%%%%%%%%%%%%%%%%%%%%%%%%%%%%%%%%%%%%%%%%%%%%%%%%%%%%%%%%%%%%%%%%%%%%%%%%%%%
%
%    THEORY
%
%%%%%%%%%%%%%%%%%%%%%%%%%%%%%%%%%%%%%%%%%%%%%%%%%%%%%%%%%%%%%%%%%%%%%%%%%%%%%%%%
\section{Estimation of gas-phase chemical abundances from optical spectroscopy}

We study a galaxy's oxygen and nitrogen abundances for several key reasons.  
These two elements are relatively abundant and emit strong lines in the optical, 
including for several ionization states in oxygen, making them relatively easy 
to observe \citep{Kewley02}.  In addition, a ratio of oxygen's lines provides a 
good estimate of the electron temperature, allowing for reliable measurements of 
a galaxy's gas-phase chemical abundances.  The following is an explanation of 
the theory and methods we employ to estimate the oxygen and nitrogen abundances 
in dwarf galaxies.

% ------------------------------------------------------------------------------
\subsection{Direct $T_e$ method}\label{sec:DirectTe}

We use the Direct $T_e$ method described in \cite{Izotov06} to estimate the 
gas-phase abundances of oxygen, nitrogen, and neon in our sample of dwarf 
galaxies, because this method is often regarded as the most accurate estimate of 
element abundances.  It can be difficult to use due to the nature of the 
[\ion{O}{3}] $\lambda$4363 auroral line \citep[for a more detailed discussion, 
see][]{Douglass17a}.  Since the strength of [\ion{O}{3}] $\lambda$4363 is 
inversely proportional to the metallicity of a galaxy, it is best suited for 
low-redshift, low-metallicity galaxies.  At metallicities \OH $\gtrsim 8.5$, 
[\ion{O}{3}] $\lambda$4363 becomes too weak to detect in the SDSS spectra.  With 
the mass-metallicity (MZ) relation by \cite{Tremonti04}, this metallicity limit 
corresponds to a maximum stellar mass of
\begin{align*}
    12 + \log \left( \frac{\text{O}}{\text{H}} \right) = 8.5 &= -1.492 + 1.847(\log M_*)\\
    &\qquad - 0.08026(\log M_*)^2\\
    \log \left( \frac{M_*}{M_\odot} \right) &= 8.696
\end{align*}
The MZ relation by \cite{Andrews13} estimates that this maximum metallicity 
translates to a maximum stellar mass of
\begin{align*}
    12 + \log \left( \frac{\text{O}}{\text{H}} \right) &= 12 + \log \left( \frac{\text{O}}{\text{H}} \right)_{asm} - \log \left( 1 + \left( \frac{M_{TO}}{M_*} \right)^\gamma \right)\\
    8.5 &= 8.798 - \log \left( 1 + \left( \frac{10^{8.901}}{M_*} \right)^{0.640} \right)\\
    M_* &= 8.138\times 10^8 M_\odot\\
    \log \left( \frac{M_*}{M_\odot} \right) &= 8.911
\end{align*}
Even though it is dangerous to compare metallicity values calculated with 
different methods, both of these upper limits on the galaxy stellar masses 
correspond to the maximum mass of a dwarf galaxy ($\log (M_*/M_\odot) \approx 
9$).  Therefore, we can expect to estimate the chemical abundances of only dwarf 
galaxies in SDSS DR7 with the Direct $T_e$ method \citep[of course, there are 
exceptions to this; see][for example]{Izotov15}.  Higher resolution spectra are 
necessary to probe higher-mass, higher-metallicity galaxies.

After solving for the temperature of the gas, we can calculate the amount of 
each element present in each of the ionization stages.  The total gas-phase 
oxygen abundance is equal to the sum of the abundances of each of the ionized 
populations:
\begin{equation}\label{eq:Oabund}
	\frac{\text{O}}{\text{H}} = \frac{\text{O}^{++}}{\text{H}^+} + \frac{\text{O}^+}{\text{H}^+}
\end{equation}

We use an ionization correction factor (ICF) to account for the missing stages 
of neon and nitrogen, since we can observe these elements in only one of their 
main ionization stages.  The total abundance for a particular element $X$ is 
\begin{equation}
    \frac{\text{X}}{\text{H}} = \sum_i ICF_i \frac{\text{X}^i}{\text{H}}
\end{equation}
For nitrogen, we employ the ICFs used in \cite{Douglass17b}.  We use the ICFs in 
\cite{Izotov06} for neon.

% ------------------------------------------------------------------------------
\subsection{O$^+$ abundance approximation}\label{sec:Oplus_approx}

\begin{figure}
    \centering
    \includegraphics[width=0.5\textwidth]{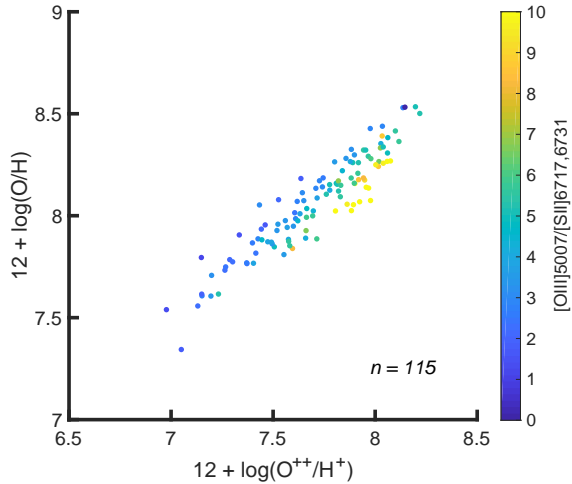}
    \caption{$12 + \log{\text{O}^{++}/\text{H}^+}$ versus \OH for star-forming 
    dwarf galaxies, as calculated with the direct method in \cite{Douglass17a}.  
    A polynomial model is fit to the data; the best fit parameters can be found 
    in the text after Eqn. \ref{eq:fit}.}
    \label{fig:lowVreal}
\end{figure}

We derive an ICF for oxygen to overcome some limitations on the observations of 
dwarf galaxies in SDSS to bolster our sample size.  Because we are studying 
dwarf galaxies, the [\ion{O}{2}] $\lambda 3727$ doublet necessary for estimating 
the abundance of O$^+$ is only available for objects within the redshift range 
$0.02 < z < 0.03$ \citep[see Sec. \ref{sec:SDSS_limits} and][for more 
details]{Douglass17a}.  To be able to study SDSS galaxies at redshifts less than 
0.02, we need to find an alternate way to estimate the abundance of O$^+$.  
While it is theoretically possible to estimate the O$^+$ abundance from the 
[\ion{O}{2}] $\lambda \lambda 7320,7331$ emission line doublet, \cite{Izotov06} 
show that the resulting abundances are not very accurate because [\ion{O}{2}] 
$\lambda \lambda 7320,7331$ is extremely weak.  Another possibility is to use an 
ICF to approximate the missing ionization stage, as outlined in Sec. 
\ref{sec:DirectTe} for the nitrogen abundance.  Because the [\ion{O}{2}] 
$\lambda 3727$ emission line is commonly available for analysis, the oxygen 
abundance can be found with Eqn. \ref{eq:Oabund}.  To construct an ICF for 
oxygen appropriate for our entire dwarf galaxy sample to include those with 
non-observed [\ion{O}{2}] $\lambda 3727$, we compare \OH to 
$12 + \log{\text{O}^{++}/\text{H}^+}$ as calculated by the Direct $T_e$ method 
for the star-forming dwarf galaxies in SDSS from \cite{Douglass17a}.  

The scatter seen in Fig. \ref{fig:lowVreal} is a result of the ionization 
parameter.  A good proxy for the ionization parameter is found with the ratio of 
two different ionization stages of the same element; [\ion{O}{3}] $\lambda 5007$ 
/ [\ion{O}{2}] $\lambda 3727$ is traditionally used \citep{Kewley02}.  However, 
if the flux of [\ion{O}{2}] $\lambda 3727$ is unknown, we must find an alternate 
proxy for the ionization ratio.  Both oxygen and sulfur are alpha elements, so 
the relative amounts of them should remain fairly constant.  Therefore, we find 
that replacing [\ion{O}{2}] $\lambda 3727$ with [\ion{S}{2}] $\lambda 6717,6731$ 
produces a suitable proxy for the ionization parameter (L. Kewley, 2018, private 
communication).

We fit a polynomial model to the data that incorporates both 
$12 + \log{\text{O}^{++}/\text{H}}$ and the ratio of [\ion{O}{3}] $\lambda 5007$ 
to [\ion{S}{2}] $\lambda \lambda 6717,6731$, defined as
\begin{equation}\label{eq:fit}
    12 + \log{\text{O}/\text{H}} = b + a_1x + a_2y + a_3xy + a_4y^2
\end{equation}
where $x = 12 + \log{\text{O}^{++}/\text{H}^+}$, 
$y = \text{[\ion{O}{3}]} \lambda 5007 / \text{[\ion{S}{2}]} \lambda \lambda 6717,6731$, 
$a_1 = 0.82 \pm 0.072$, $a_2 = -0.3\pm 0.15$, $a_3 = 0.03\pm 0.019$, 
$a_4 = 0.0006\pm 0.00032$, and $b = 1.9 \pm 0.55$ for star-forming dwarf 
galaxies.

For those star-forming dwarf galaxies for which [\ion{O}{2}] $\lambda$3727 is 
not observed in the SDSS spectra, we use the coefficients for the star-forming 
dwarf galaxies with Eqn. \ref{eq:fit} to calculate the total oxygen abundance in 
the galaxy based on the amount of doubly ionized oxygen found in Eqn. 4 of 
\cite{Douglass17a}.

%%%%%%%%%%%%%%%%%%%%%%%%%%%%%%%%%%%%%%%%%%%%%%%%%%%%%%%%%%%%%%%%%%%%%%%%%%%%%%%%
%
%    DATA
%
%%%%%%%%%%%%%%%%%%%%%%%%%%%%%%%%%%%%%%%%%%%%%%%%%%%%%%%%%%%%%%%%%%%%%%%%%%%%%%%%
\section{SDSS data and galaxy selection}

The SDSS Data Release 7 \citep[DR7;][]{Abazajian09} is a wide-field multiband 
imaging and spectroscopic survey employing a drift scanning technique to map 
approximately one quarter of the northern sky.  Photometric data in the 
five-band SDSS system --- $u$, $g$, $r$, $i$, and $z$ --- are taken with a 
dedicated 2.5~m telescope at the Apache Point Observatory in New Mexico 
\citep{Fukugita96, Gunn98}.  Follow-up spectroscopic analysis is performed on 
galaxies with a Petrosian $r$-band magnitude $m_r < 17.77$ \citep{Lupton01, 
Strauss02}.  The spectra are taken using two double fiber-fed spectrometers and 
fiber plug plates with a minimum fiber separation of 55"; the observed 
wavelength range is 3800--9200\AA ~with a resolution 
$\lambda / \Delta \lambda \sim$1800 \citep{Blanton03}.  We use emission-line 
flux data from the MPA-JHU value-added catalog,\footnote{Available at 
\url{http://www.mpa-garching.mpg.de/SDSS/DR7/}} which is based on the SDSS DR7 
sample of galaxies; as described at 
\url{http://www.sdss.org/dr12/spectro/galaxy_mpajhu/} and in \cite{Tremonti04}, 
this catalog's emission line analysis includes a subtraction of the best-fitting 
stellar population model of the continuum, thus accounting for the underlying 
stellar absorption features.  All flux values have been corrected for dust 
reddening with the \cite{Cardelli89} extinction curve as implemented in pyNeb 
\citep{Luridiana15}; we assume the theoretical ratio H$\alpha$/H$\beta = 2.86$ 
at 10,000 K and 100 cm$^{-3}$ \citep{Osterbrock89}.  As shown in 
\cite{LopezSanchez15}, the magnitude of this ratio is mildly dependent on the 
electron temperature.  This ratio varies by $\sim$7\% over the range of 
temperatures seen in this sample, which translates to a difference in the 
metallicity of less than 0.003 dex, well within the uncertainty in the 
metallicity estimates due to the fluxes.

We use the stellar mass estimates from the NASA-Sloan Atlas \citep{Blanton11}.  
The \ion{H}{1} mass estimates are from the 70\% complete ALFALFA catalog 
$\alpha.70$ \citep{Giovanelli05}; \ion{H}{1} detections were matched to the SDSS 
galaxies by locating the nearest optical counterpart identified in the 
$\alpha.70$ catalog within 1 arcmin.  Absolute magnitudes, colors, and all other 
additional data are from the KIAS value-added galaxy catalog 
\citep{Blanton05,Choi10}.  Galaxy colors are rest-frame colors that have been 
$K$-corrected to a redshift of 0.1; they are corrected for galactic extinction 
and calculated with model magnitudes.  All galaxies have been visually inspected 
to remove any galaxy fragments or duplicates.

%-------------------------------------------------------------------------------
\subsection{Spectroscopic selection}\label{sec:SDSS_limits}

We employ most of the same requirements for our sample as in 
\cite{Douglass17a,Douglass17b}.  All galaxies must have
\begin{enumerate}
    \item{$M_r > -17$ (dwarf galaxies);}
    \item{a minimum $5\sigma$ detection of H$\beta$;}
    \item{a minimum $1\sigma$ detection of [\ion{O}{3}] $\lambda 4363$;}
    \item{a flux $> 0$ for [\ion{O}{2}] $\lambda 3727$, [\ion{O}{3}] $\lambda \lambda 4959,5007$, and [\ion{N}{2}] $\lambda \lambda 6548,6584$;}
    \item{$T(\text{[\ion{O}{3}]}) < 2\times 10^4 \text{ K}$;}
    \item{a star-forming BPT classification by \cite{Brinchmann04}.}
\end{enumerate}

For those galaxies with a redshift $z \gtrsim 0.02$, we use the 
\texttt{oii\_flux} value from the MPA-JHU catalog in place of their [\ion{O}{2}] 
$\lambda \lambda 3726,3729$ flux measurement.  See \cite{Douglass17a} for 
further details on each of these requirements.

We require the estimated electron temperature 
$T(\text{[\ion{O}{3}]}) < 2\times 10^4 \text{ K}$ for a few reasons.  
\cite{Nicholls14c} explain that temperatures above this value estimated from the 
direct method as prescribed by \cite{Izotov06} are not reliable because the data 
used to derive these equations do not extend beyond $2\times 10^4 \text{ K}$.  
Physically, one does not typically find \ion{H}{2} regions with temperatures 
higher than $2\times 10^4 \text{ K}$ because the equilibrium temperature between 
photoelectric heating and cooling due to recombination, free-free emission, and 
collisionally excited line radiation is often below this temperature.  However, 
some extremely metal-deficient galaxies are found with temperatures higher than 
$2\times 10^4 \text{ K}$ \citep[for example]{Guseva17}; by eliminating these 
metal-poor galaxies, we are introducing a sample bias.  We examine the 
significance of this bias in Sec. \ref{sec:systematics}.

%-------------------------------------------------------------------------------
\subsection{Void classification}

The large-scale environment of the galaxies is determined using the void catalog 
compiled by \cite{Pan12}, which was constructed using galaxies in the SDSS DR7 
catalog.  The VoidFinder algorithm of \cite{Hoyle02} \citep[based on the 
algorithm described by][]{ElAd97} removes all isolated galaxies (defined as 
having the third nearest neighbor more than 7 $h^{-1}$Mpc away), using only 
galaxies with absolute magnitudes $M_r < -20$.  After applying a grid to the 
remaining galaxies, spheres are grown from all cells containing no galaxies 
until it encounters four galaxies on its surface.  A sphere must have a minimum 
10 $h^{-1}$Mpc radius to be classified as a void (or part of one).  If two 
spheres overlap by more than 50\%, they are considered part of the same void.  
See \cite{Hoyle02} for a more detailed description of the VoidFinder algorithm.  
Those galaxies that fall within these void spheres are classified as voids.  
Galaxies that lie outside the spheres are classified as wall galaxies.  Because 
we cannot identify the center of any void within 5 $h^{-1}$Mpc of the edge of 
the survey, we classify those galaxies within this border region as 
``Uncertain.''

Of the $\sim$800,000 galaxies with spectra available in SDSS DR7, 9519 are 
dwarf galaxies.  Applying the spectroscopic cuts, our sample includes 993 void 
dwarf galaxies and 759 wall dwarf galaxies.

%%%%%%%%%%%%%%%%%%%%%%%%%%%%%%%%%%%%%%%%%%%%%%%%%%%%%%%%%%%%%%%%%%%%%%%%%%%%%%%%
%
%    ANALYSIS & RESULTS
%
%%%%%%%%%%%%%%%%%%%%%%%%%%%%%%%%%%%%%%%%%%%%%%%%%%%%%%%%%%%%%%%%%%%%%%%%%%%%%%%%
\section{Abundance analysis and results}

Our primary objective is to perform a relative measurement of gas-phase 
abundances of dwarf galaxies to discern how their chemical evolution is affected 
by the large-scale environment.

All line ratios listed are ratios of the emission-line fluxes.  Galaxies with 
low metallicities have $Z =$ \OH $< 7.6$ \citep{Pustilnik06}; galaxies with high 
metallicities have $Z > 8.2$ \citep{Pilyugin06}.  The solar metallicity 
$Z_\odot = 8.69\pm 0.05$ \citep{Asplund09}.

%-------------------------------------------------------------------------------
\subsection{Dwarf galaxy abundances}

% Results table (machine-readable)
\floattable
\begin{deluxetable}{ccccccccccccccc}
\tablewidth{0pt}
\tablehead{\colhead{Index\tablenotemark{a}} & \colhead{R.A.} & \colhead{Decl.} & \colhead{Redshift} & \multicolumn{2}{c}{$12 + \log \left( \frac{\text{O}}{\text{H}} \right)$} & \multicolumn{2}{c}{$12 + \log \left( \frac{\text{N}}{\text{H}} \right)$} & \multicolumn{2}{c}{$12 + \log \left( \frac{\text{Ne}}{\text{H}} \right)$} & \multicolumn{2}{c}{$\log \left( \frac{\text{N}}{\text{O}} \right)$} & \multicolumn{2}{c}{$\log \left( \frac{\text{Ne}}{\text{O}} \right)$} & \colhead{Void/Wall}}
\tablecaption{Analyzed dwarf galaxies\label{tab:Results}}
%\tablehead{\colhead{index} & \colhead{ra} & \colhead{dec} & \colhead{redshift} & \colhead{Z12logOH} & \colhead{Zerr} & \colhead{N12logNH} & \colhead{NHerr} & \colhead{Ne12logNeH} & \colhead{NeHerr} & \colhead{logNO} & \colhead{NOerr} & \colhead{logNeO} & \colhead{NeOerr} & \colhead{vflag}}
\startdata
40726 & \RA{11}{25}{52}{.10} & -\dec{00}{39}{41}{.76} & 0.0187 & 8.15 & $\pm$0.09 & 6.72 & $\pm$0.06 & 7.44 & $\pm$0.12 & -1.43 & $\pm$0.11 & -0.71 & $\pm$0.15 & Wall \\
41257 & \RA{12}{41}{12}{.41} & -\dec{00}{45}{24}{.55} & 0.0113 & 8.16 & $\pm$0.24 & 6.96 & $\pm$0.16 & 7.35 & $\pm$0.31 & -1.20 & $\pm$0.29 & -0.81 & $\pm$0.40 & Wall \\
42296 & \RA{14}{39}{50}{.03} & -\dec{00}{42}{22}{.85} & 0.0060 & 7.93 & $\pm$0.12 & 6.72 & $\pm$0.08 & 7.30 & $\pm$0.16 & -1.21 & $\pm$0.14 & -0.63 & $\pm$0.20 & Wall \\
42829 & \RA{15}{40}{18}{.50} & -\dec{00}{48}{45}{.04} & 0.0125 & 8.10 & $\pm$0.36 & 6.92 & $\pm$0.25 & 7.41 & $\pm$0.48 & -1.18 & $\pm$0.43 & -0.68 & $\pm$0.60 & Wall \\
44522 & \RA{11}{47}{00}{.73} & -\dec{00}{17}{39}{.22} & 0.0049 & 8.10 & $\pm$0.25 & 6.91 & $\pm$0.17 & 7.53 & $\pm$0.33 & -1.19 & $\pm$0.30 & -0.58 & $\pm$0.42 & Wall
\enddata
\tablecomments{Five of the 1561 dwarf galaxies analyzed from SDSS DR7.  The flux values for all required emission lines can be found in the MPA-JHU value-added catalog.  Abundance values are calculated using the direct $T_e$ method and the O$^+$ approximation defined in Sec. \ref{sec:Oplus_approx}, with error estimates via a Monte-Carlo method.  The void catalog of \cite{Pan12} is used to classify the galaxies as either Void or Wall.  If a galaxy is located too close to the boundary of the SDSS survey to identify whether or not it is inside a void, it is labeled as uncertain.  Table \ref{tab:Results} is published in its entirety online in a machine-readable format.  A portion is shown here for guidance regarding its form and content.}
\tablenotetext{a}{KIAS-VAGC galaxy index number}
\end{deluxetable}

The abundances estimated using the direct $T_e$ method for our dwarf galaxy 
sample are listed in Table \ref{tab:Results}.  Also included are other 
significant characteristics and identification for the galaxies, including their 
large-scale environmental classification.

\subsubsection{Oxygen, nitrogen, and neon abundances}

\begin{figure*}
    \centering
    \includegraphics[width=0.49\textwidth]{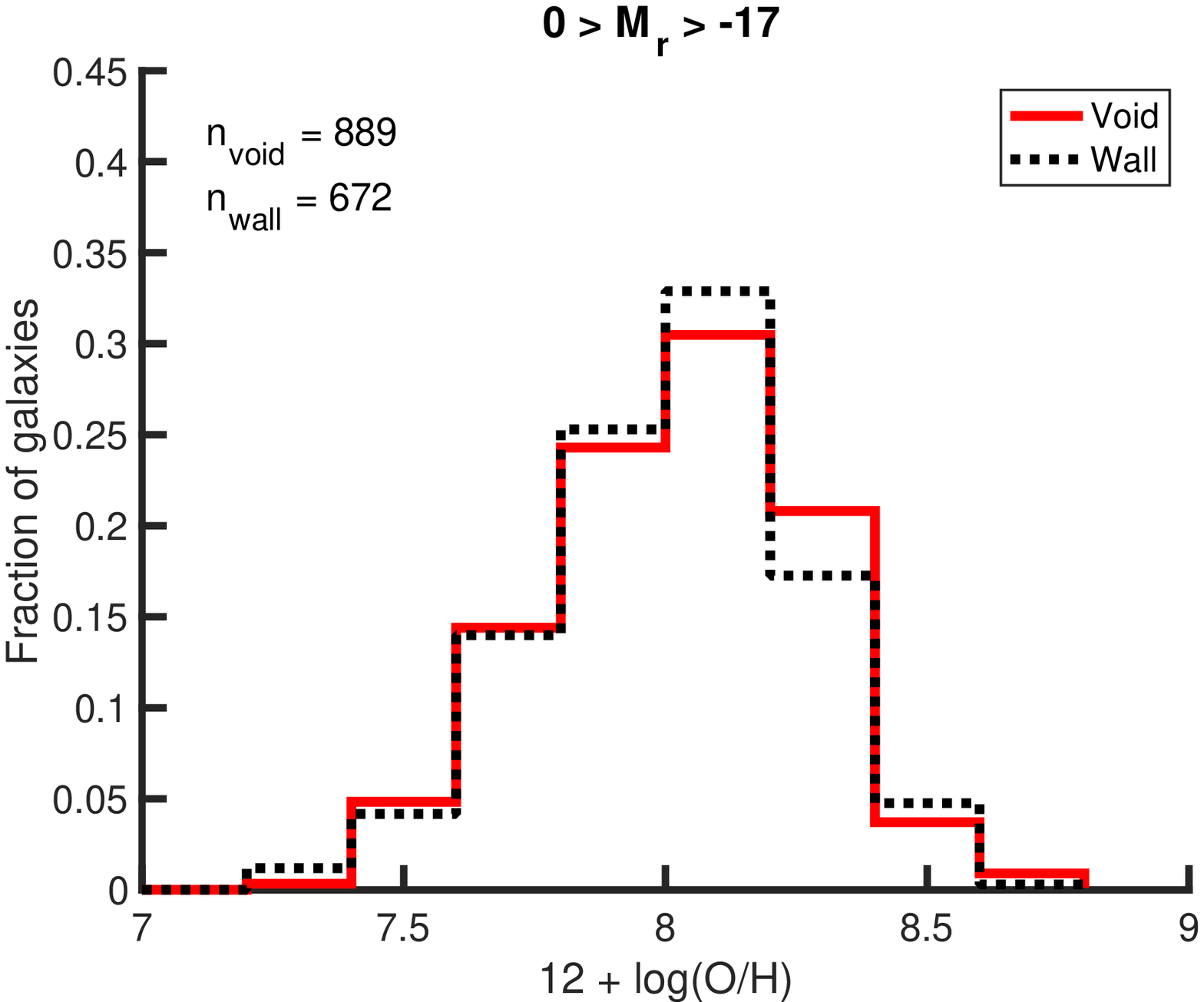}
    \includegraphics[width=0.49\textwidth]{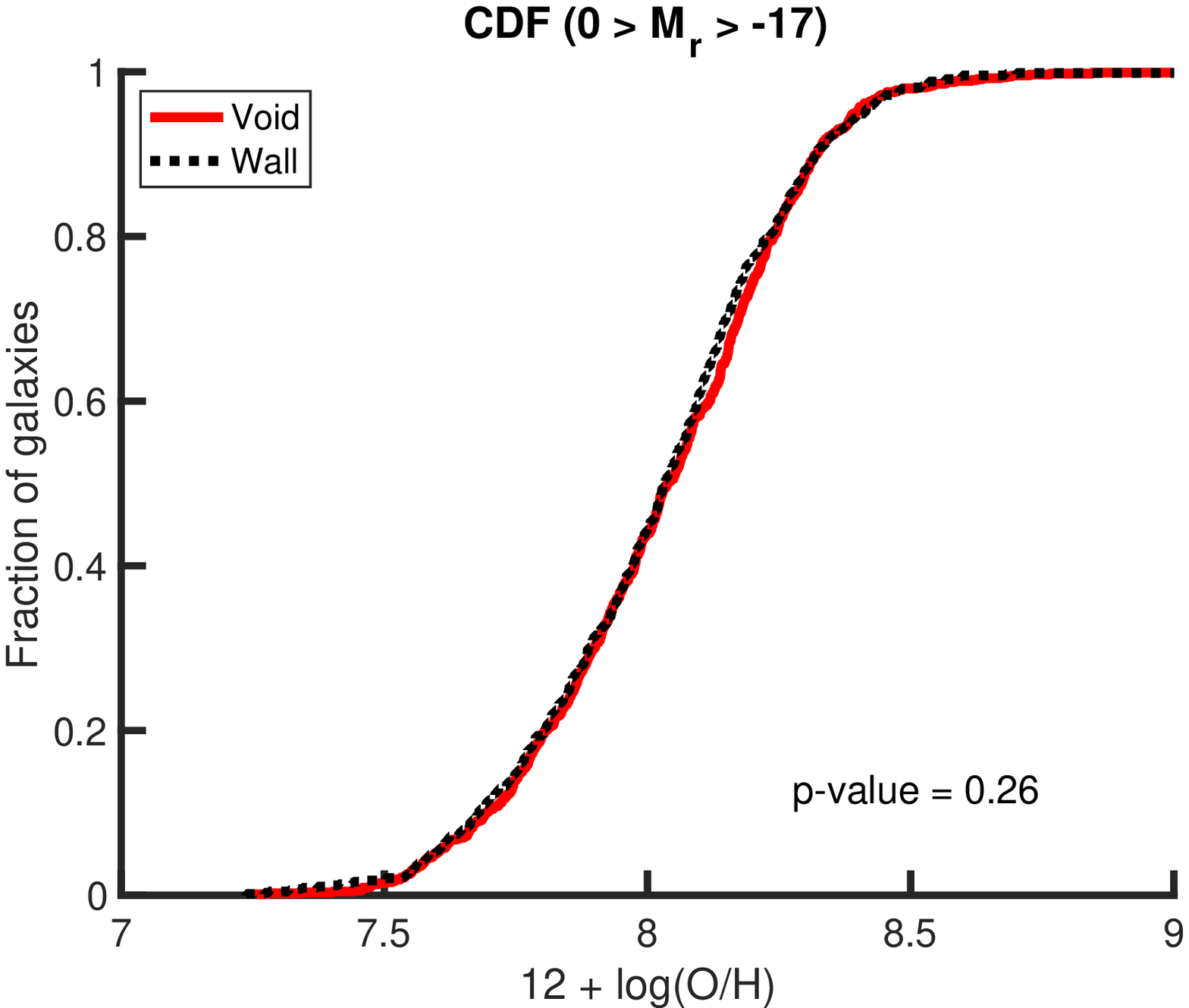}
    \caption{Gas-phase metallicity of void dwarf (red solid line) and wall 
    dwarf (black dashed line) galaxies.  A two-sample K-S test of the two data 
    sets results in an asymptotic $p$-value of 0.259, indicating a 26\% 
    probability that a test statistic greater than the observed value of 0.051 
    will be seen if the void sample is drawn from the wall sample.  This is 
    reflected visually, as there appears to be very little difference between 
    the metallicity distributions of the void and wall dwarf galaxy sample 
    populations.}
    \label{fig:met1sig}
\end{figure*}

\begin{figure*}
    \centering
    \includegraphics[width=0.49\textwidth]{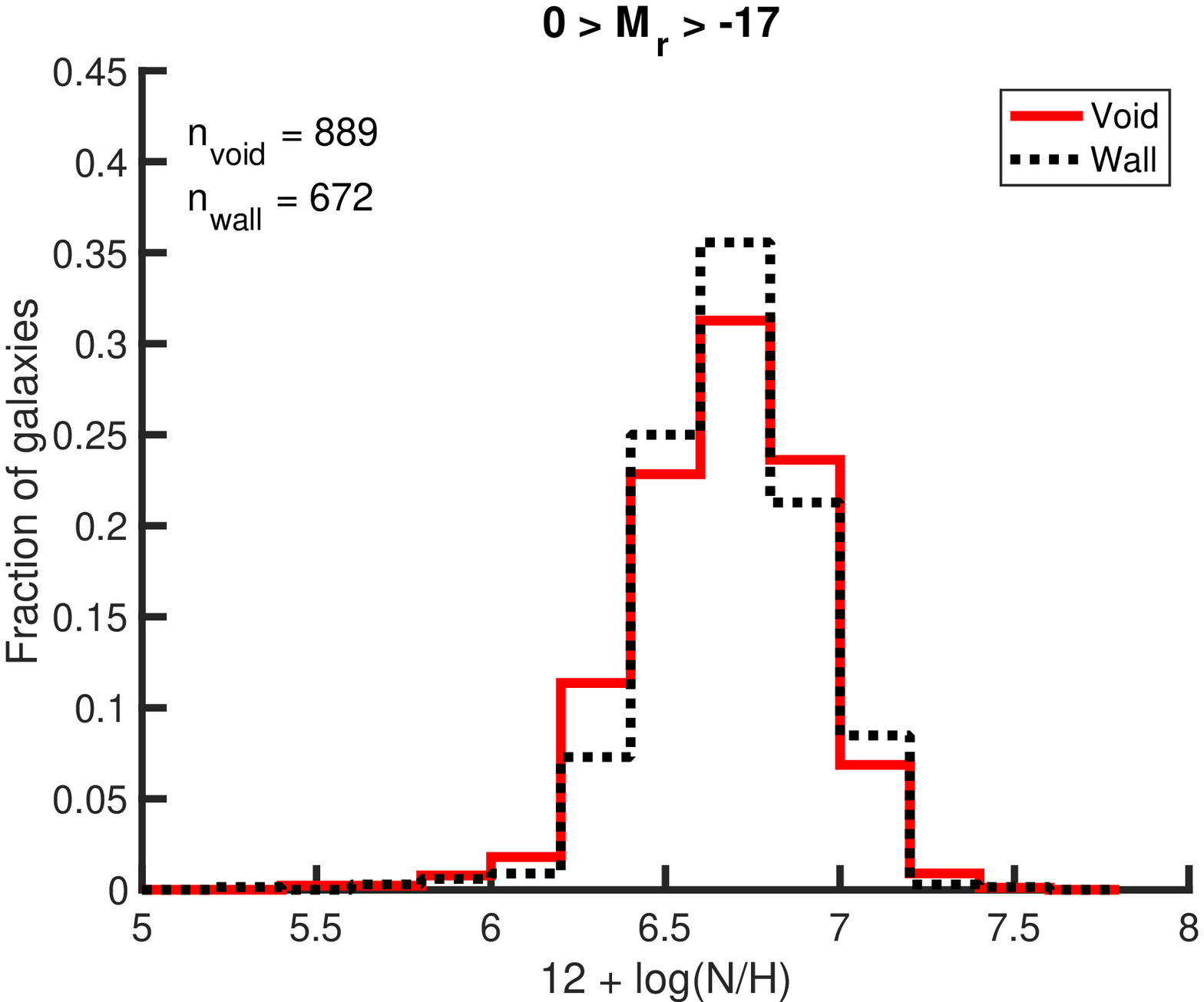}
    \includegraphics[width=0.49\textwidth]{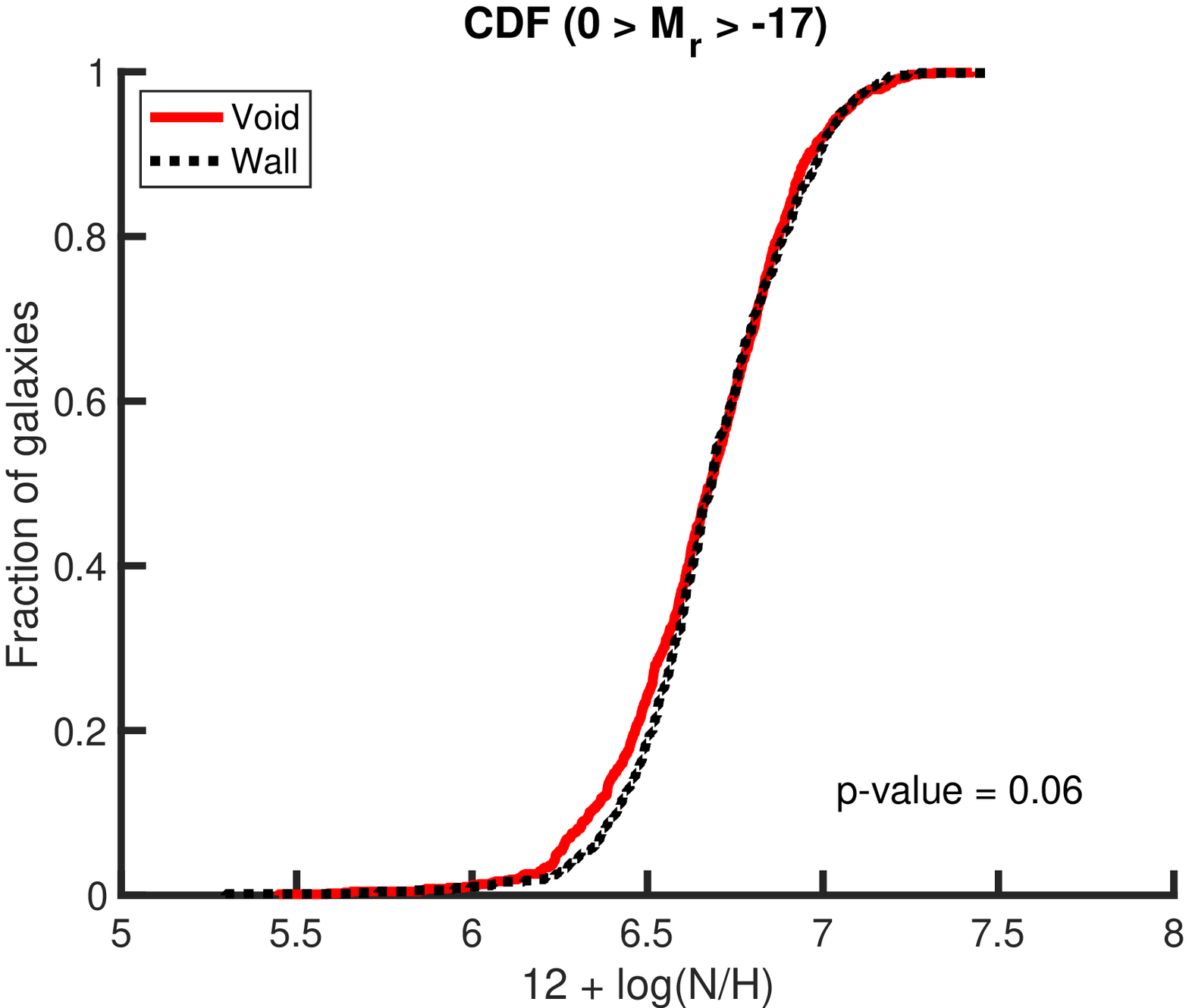}
    \caption{Abundance of nitrogen relative to hydrogen of void dwarf (red solid 
    line) and wall dwarf (black dashed line) galaxies.  A two-sample K-S test of 
    the two data sets results in an asymptotic $p$-value of 0.058, indicating a 
    5.8\% probability that a test statistic greater than the observed value of 
    0.068 will be seen, if the void sample is drawn from the wall sample.  This 
    is reflected visually, as it appears that there are slightly more void dwarf 
    galaxies with low nitrogen abundances than in the wall sample.}
    \label{fig:N_1sig}
\end{figure*}

\begin{figure*}
    \centering
    \includegraphics[width=0.49\textwidth]{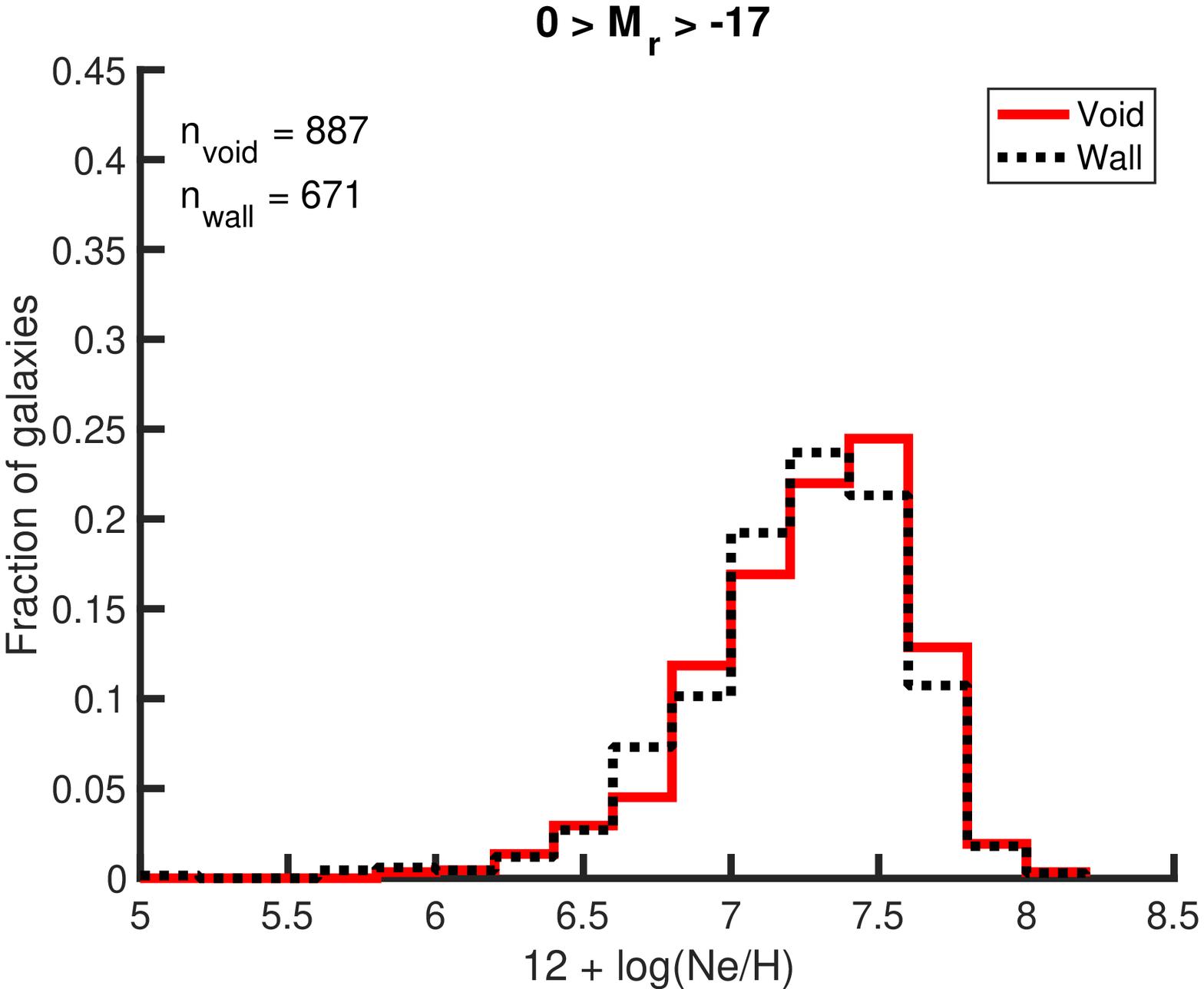}
    \includegraphics[width=0.49\textwidth]{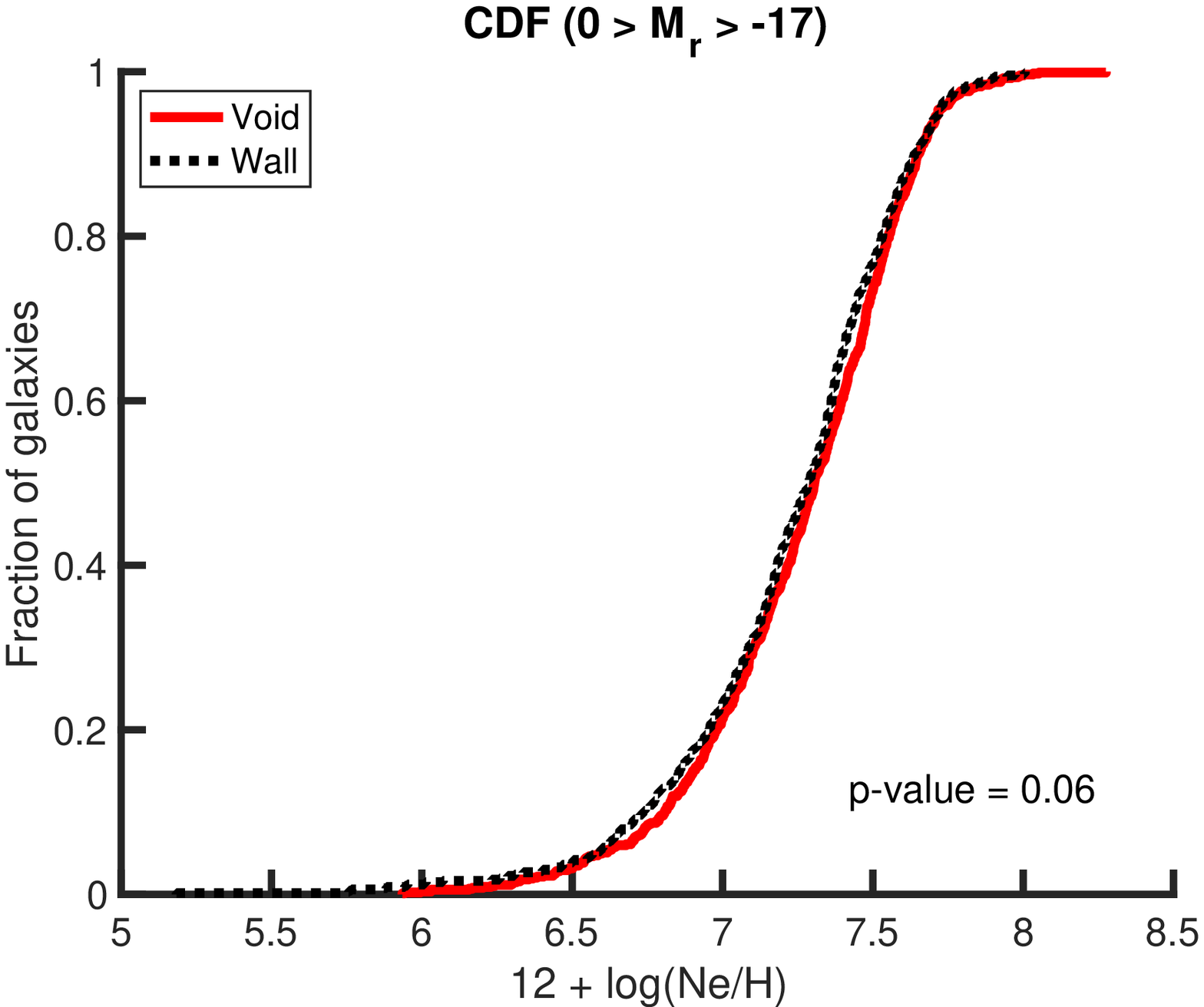}
    \caption{Abundance of neon relative to hydrogen of void dwarf (red solid 
    line) and wall dwarf (black dashed line) galaxies.  A two-sample K-S test of 
    the two data sets results in an asymptotic $p$-value of 0.055, indicating a 
    5.5\% probability that a test statistic greater than the observed value of 
    0.068 will be seen if the void sample is drawn from the wall sample.  This 
    is reflected visually, as the void dwarf galaxies appear to have higher 
    values of the Ne/H ratio than the wall dwarf galaxies.}
    \label{fig:Ne_1sig}
\end{figure*}

The distributions of oxygen, nitrogen, and neon abundances for dwarf galaxies as 
a function of large-scale environment are shown in Figures 
\ref{fig:met1sig}--\ref{fig:Ne_1sig}, respectively.  All three histograms show a 
slight shift between voids and walls in the chemical abundances of dwarf 
galaxies.  A two-sample Kolmogorov--Smirnov (K-S) test quantifies this 
observation: a test statistic of 0.05 for oxygen, 0.07 for nitrogen, and 0.07 
for neon are produced, corresponding to a probability of 26\%, 5.8\%, and 5.5\%, 
respectively, that a test statistic greater than or equal that observed will be 
measured if the void sample was drawn from the wall sample.  The cumulative 
distribution function (CDF) for each of these elements can be seen on the right 
in Figures \ref{fig:met1sig}--\ref{fig:Ne_1sig}; they show that void dwarf 
galaxies have roughly the same oxygen abundances, slightly higher neon 
abundances, and slightly lower nitrogen abundances than dwarf galaxies in more 
dense regions.  The K-S test quantifies the visual interpretation of these 
figures that the distribution of oxygen is not significantly different for 
star-forming dwarf galaxies in voids and walls, while the distributions of 
nitrogen and neon abundances do differ slightly.

The average and median values of the dwarf galaxy abundances also indicate 
either a slight shift or no shift as a result of the large-scale environment.  
The average oxygen abundance for void dwarf galaxies is $8.03\pm 0.009$ and the 
median is 8.04, and the average oxygen abundance for wall dwarf galaxies is 
$8.02\pm 0.010$ with a median value of 8.03.  This implies that there is no 
significant difference in the oxygen abundances between void and wall dwarf 
galaxies (average shift of $0.01\pm 0.013$; median shift of 0.01).  The average 
nitrogen abundance for void dwarf galaxies is only $6.66\pm 0.006$ with a median 
of 6.68, while the wall dwarf galaxies have an average nitrogen abundance of 
$6.69\pm 0.007$ and a median of 6.68.  The void dwarf galaxies have slightly 
lower nitrogen abundances by about 5\% (an average shift of $0.02\pm 0.009$ and 
a median shift of 0.00) relative to wall dwarf galaxies.  In contrast, the 
average neon abundance for void dwarf galaxies is $7.26\pm 0.012$ and the median 
is 7.30, while the average neon abundance for wall dwarf galaxies is 
$7.22\pm 0.014$ with a median value of 7.28.  The void dwarf galaxies have 
slightly higher neon abundances by about 10\% (an average shift of 
$0.04\pm 0.019$; a median shift of 0.02) relative to the wall dwarf galaxies.  A 
tabular version of this analysis can be found in Table \ref{tab:stats}.

\subsubsection{Ratio of nitrogen, neon to oxygen}

\begin{figure*}
    \centering
    \includegraphics[width=0.49\textwidth]{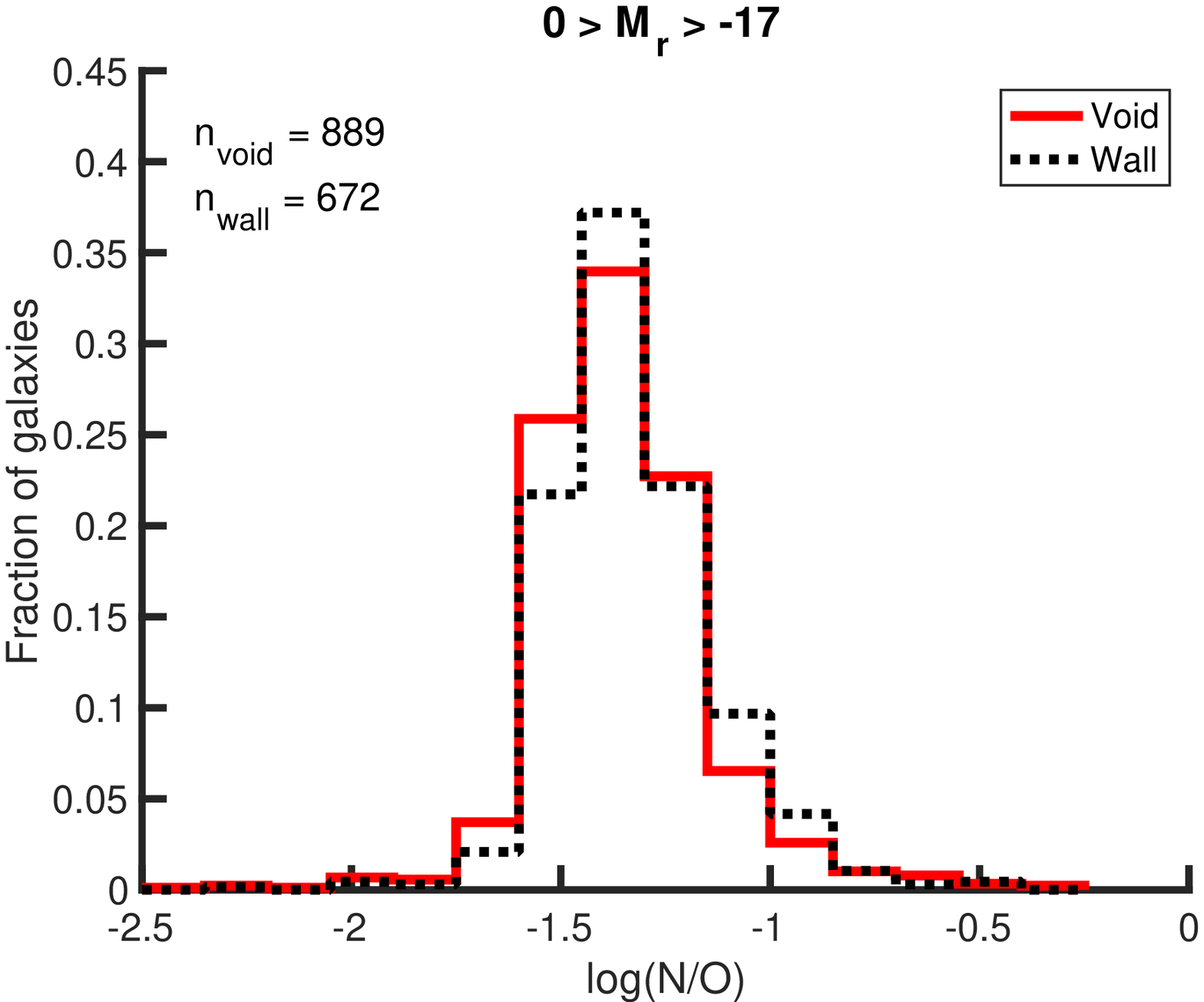}
    \includegraphics[width=0.49\textwidth]{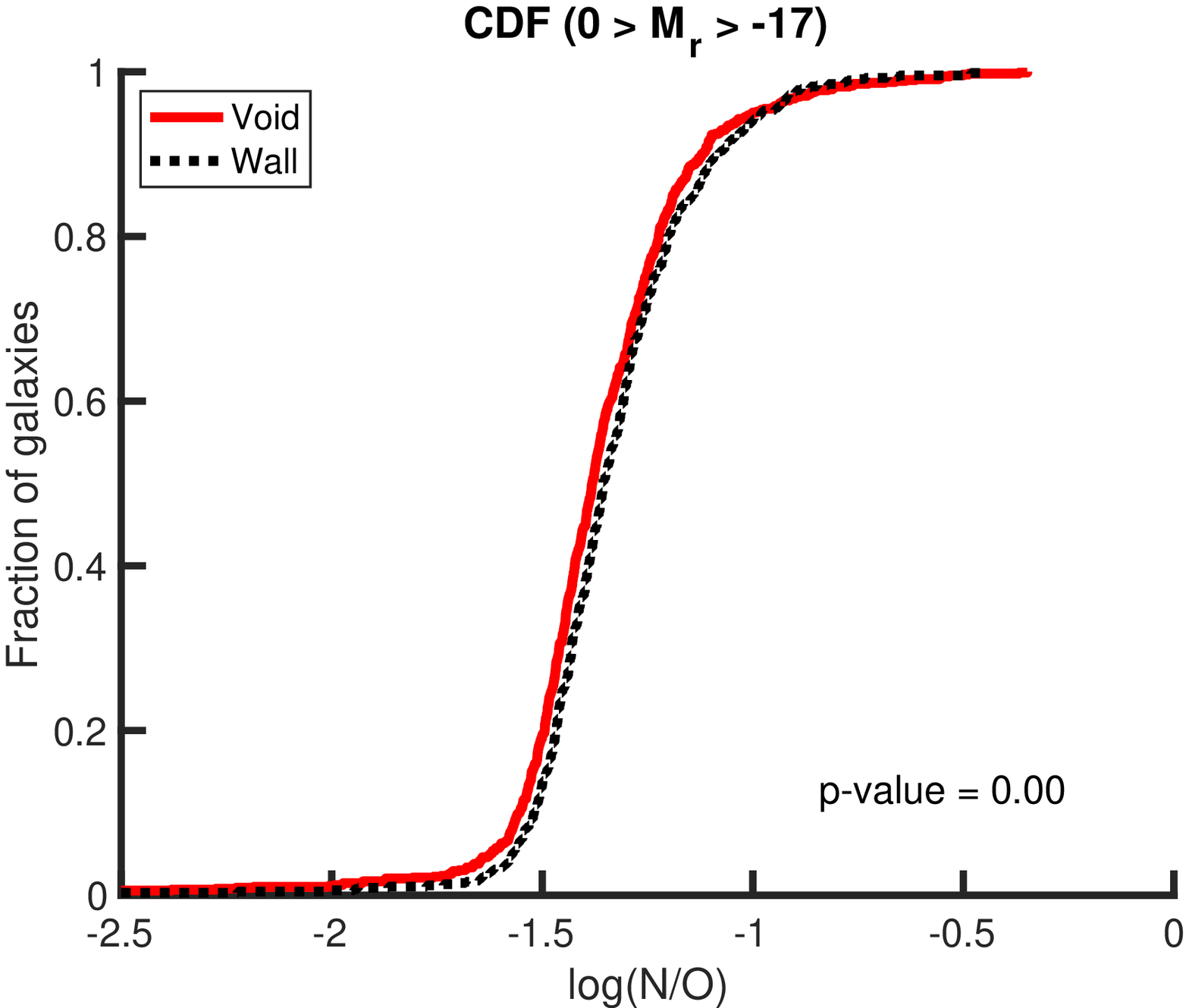}
    \caption{Ratio of nitrogen to oxygen of void dwarf (red solid line) and wall 
    dwarf (black dashed line) galaxies.  A two-sample K-S test of the two data 
    sets results in an asymptotic $p$-value of $3.6\times 10^{-3}$, indicating 
    only a 0.4\% probability that a test statistic greater than the observed 
    value of 0.09 will be seen.  This is reflected visually, as there is a shift 
    in the N/O ratio between the two populations of dwarf galaxies --- the void 
    galaxies have a lower value of N/O than the wall galaxies.}
    \label{fig:NOratio}
\end{figure*}

\begin{figure*}
    \centering
    \includegraphics[width=0.49\textwidth]{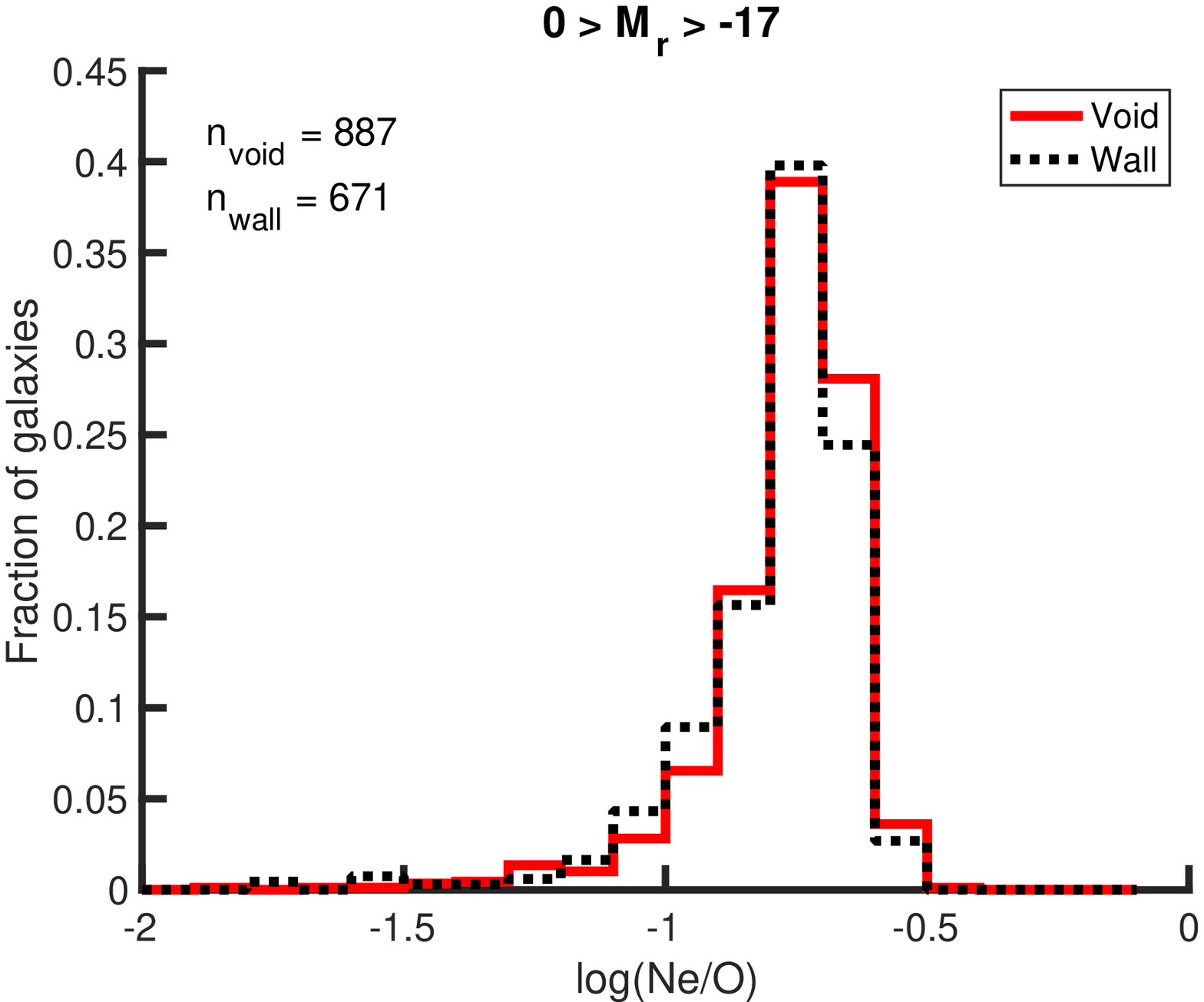}
    \includegraphics[width=0.49\textwidth]{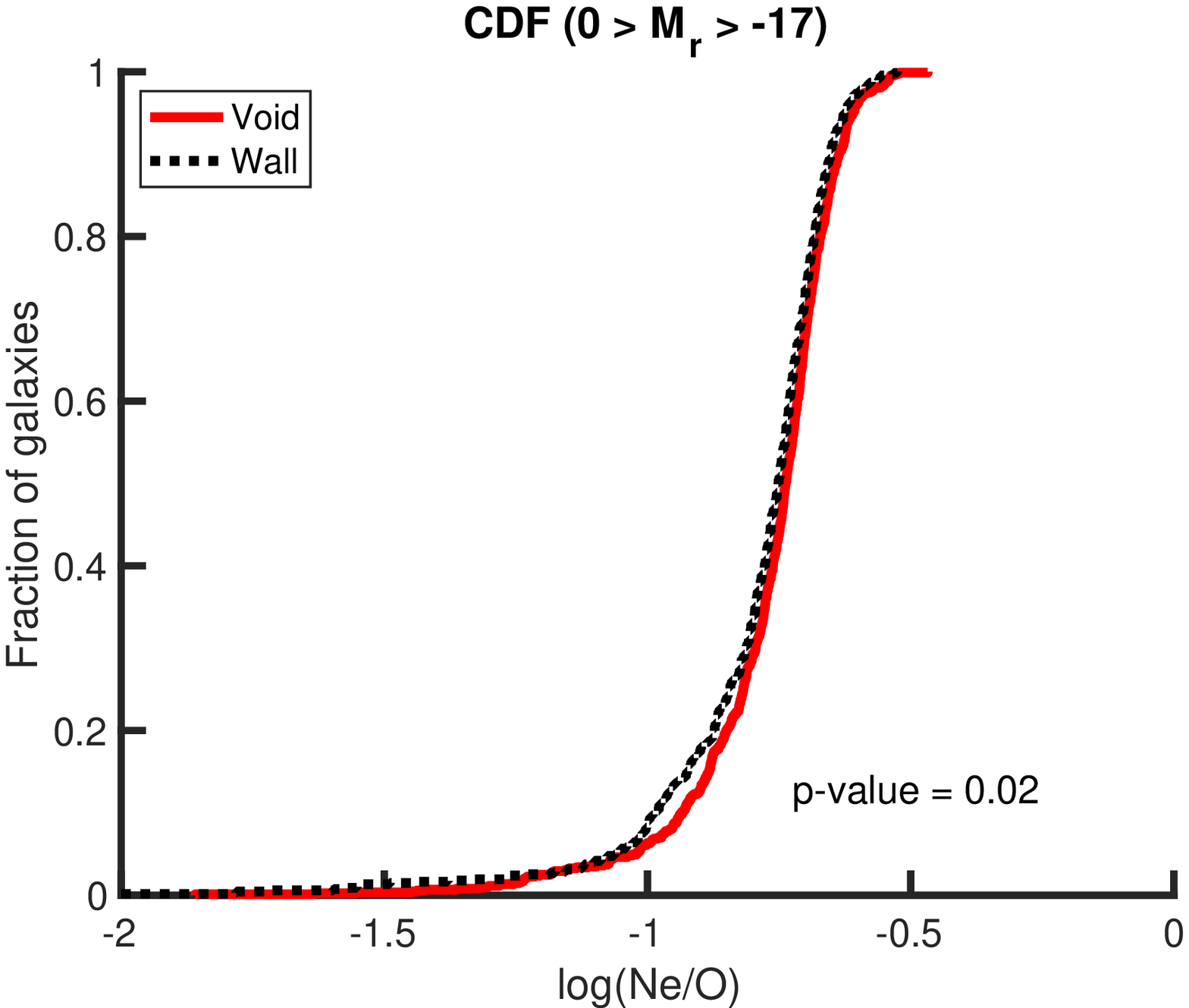}
    \caption{Ratio of neon to oxygen of void dwarf (red solid line) and wall 
    dwarf (black dashed line) galaxies.  A two-sample K-S test of the two data 
    sets results in an asymptotic $p$-value of 0.02, indicating only a 2\% 
    probability that a test statistic greater than the observed value of 0.077 
    will be seen.  This is reflected visually, as there is a slight shift in the 
    Ne/O ratio between the two populations of dwarf galaxies --- the void 
    galaxies have a slightly higher value of Ne/O than the wall galaxies.}
    \label{fig:NeOratio}
\end{figure*}

The ratio of nitrogen to oxygen is also important to investigate, as it 
communicates the nucleosynthesis history of the galaxies.  As seen in Fig. 
\ref{fig:NOratio}, the N/O abundance ratio indicates a slight large-scale 
environmental influence on the chemical evolution of dwarf galaxies --- void 
dwarf galaxies have slightly lower N/O ratios than dwarf galaxies in denser 
regions.  This difference is quantified in the K-S test: the test returned a 
probability of only 0.4\% that a test statistic greater than or equal to 0.09 
will be measured if the void sample was drawn from the wall sample.  The 
distribution of N/O abundance ratios for void dwarf galaxies is lower by about 
7\% (an average shift of $0.03\pm 0.016$ and a median shift of 0.03) relative to 
the distribution of N/O ratios in wall dwarf galaxies.

In contrast, the ratio of neon to oxygen serves as a diagnostic of the accuracy 
of the abundance estimates.  Because neon and oxygen are both synthesized via 
$\alpha$ processes in the same mass stars, they should always be produced in the 
same fixed ratio, assuming that the ratio of the IMF-averaged Ne and O yields 
does not depend on stellar metallicities.  Even if the void environment 
influences the chemical evolution of dwarf galaxies, there should exist only a 
minor (if any) shift in the Ne/O ratio.  As can be seen in Fig. 
\ref{fig:NeOratio}, there is very little difference in the distributions of the 
Ne/O ratio between the two populations.  The K-S test quantifies this 
difference: the test returns a probability of 2\% that a test statistic greater 
than or equal to 0.077 will be measured if the void sample was drawn from the 
wall sample.  Unlike the shifts measured in the distributions of nitrogen, neon, 
and N/O, the average shift of $0.02\pm 0.023$ in the Ne/O abundance ratio is 
within the uncertainty of the shift.  As physically expected, the void 
environment has almost no influence on the relative synthesis of neon and oxygen 
in star-forming dwarf galaxies.

% Statistics table
\floattable
\begin{deluxetable}{ccccccc}
    \tablewidth{0pt}
    \tablecolumns{7}
    \tablecaption{Abundance statistics\label{tab:stats}}
    \tablehead{\colhead{Environment} & \colhead{Average} & \colhead{Median} & \colhead{Average Shift\tablenotemark{a}} & \colhead{Median Shift\tablenotemark{a}} & \colhead{$p$-value} & \colhead{K-S Test Statistic}}
    \startdata
        % Oxygen %%%%%%%%%%%%%%%%%%%%%%%%%%%%%%%%%%%%%%%%%%%%%%%%%%%%%%%%%%%%%%%
        \cutinhead{\OH}
        Void & $8.03\pm 0.009$ & 8.04 & \multirow{2}{*}{$-0.01\pm 0.013$} & \multirow{2}{*}{-0.01} & \multirow{2}{*}{0.259} & \multirow{2}{*}{0.0513}\\
        Wall & $8.02\pm 0.010$ & 8.03 & & & & \\
        % Nitrogen %%%%%%%%%%%%%%%%%%%%%%%%%%%%%%%%%%%%%%%%%%%%%%%%%%%%%%%%%%%%
        \cutinhead{\NH}
        Void & $6.66\pm 0.006$ & 6.68 & \multirow{2}{*}{$0.02\pm 0.009$} & \multirow{2}{*}{0.00} & \multirow{2}{*}{0.0575} & \multirow{2}{*}{0.0677}\\
        Wall & $6.69\pm 0.007$ & 6.68 & & & & \\
        % Neon %%%%%%%%%%%%%%%%%%%%%%%%%%%%%%%%%%%%%%%%%%%%%%%%%%%%%%%%%%%%%%%%
        \cutinhead{\NeH}
        Void & $7.26\pm 0.012$ & 7.30 & \multirow{2}{*}{$-0.04\pm 0.019$} & \multirow{2}{*}{-0.02} & \multirow{2}{*}{0.0554} & \multirow{2}{*}{0.0681}\\
        Wall & $7.22\pm 0.014$ & 7.28 & & & & \\
        % N/O %%%%%%%%%%%%%%%%%%%%%%%%%%%%%%%%%%%%%%%%%%%%%%%%%%%%%%%%%%%%%%%%%
        \cutinhead{\NO}
        Void & $-1.36\pm 0.011$ & -1.38 & \multirow{2}{*}{$0.03\pm 0.016$} & \multirow{2}{*}{0.03} & \multirow{2}{*}{0.00364} & \multirow{2}{*}{0.0902}\\
        Wall & $-1.33\pm 0.012$ & -1.36 & & & & \\
        % Ne/O %%%%%%%%%%%%%%%%%%%%%%%%%%%%%%%%%%%%%%%%%%%%%%%%%%%%%%%%%%%%%%%%
        \cutinhead{\NeO}
        Void & $-0.77\pm 0.015$ & -0.74 & \multirow{2}{*}{$-0.02\pm 0.023$} & \multirow{2}{*}{-0.01} & \multirow{2}{*}{0.0199} & \multirow{2}{*}{0.0772}\\
        Wall & $-0.79\pm 0.018$ & -0.75 & & & & \\
    \enddata
    \tablecomments{Statistics on the gas-phase oxygen, nitrogen, neon, nitrogen 
    relative to oxygen, and neon relative to oxygen abundances in dwarf void and 
    wall galaxies.  Combined with the histograms in Figures 
    \ref{fig:met1sig}--\ref{fig:NeOratio}, these results indicate an influence 
    on the chemical evolution of galaxies by the large-scale environment, 
    especially on the relative abundance of nitrogen to oxygen.  Void galaxies 
    have slightly higher neon abundances and slightly lower nitrogen abundances 
    and N/O ratios than wall galaxies.  There is no significant shift in the 
    oxygen abundance or the Ne/O ratio between void and wall dwarf galaxies.}
    \tablenotetext{a}{Wall/Void (Positive shifts indicate that the wall values 
    are greater than the void values; negative shifts indicate that the void 
    values are greater than the wall values.)}
\end{deluxetable}

%-------------------------------------------------------------------------------
\subsection{Estimation of uncertainties}

Uncertainties in the computed abundances are estimated using a Monte-Carlo 
method.  Using the measured line fluxes and scaled uncertainty estimates, we 
calculate 100,000 abundance estimates.  For each abundance estimate, a new 
positive ``fake'' line flux is drawn from a normal distribution.  We use the 
standard deviation in these sets of 100,000 calculated abundance values for the 
error in our abundance estimate.  See \cite{Douglass17a} for a more in-depth 
description of this process.

%-------------------------------------------------------------------------------
\subsection{Sources of systematic error}\label{sec:systematics}

Many physical properties of galaxies exhibit a radial dependence \citep{Bell00}.  
As a result, abundance estimates can depend on where the spectroscopic fiber is 
placed on the galaxy.  The estimated abundance will not necessarily be 
representative of a global abundance value if not all of the galaxy's light is 
contained within the fiber.  \cite{Belfiore17} show that both the metallicity 
and N/O ratio gradients are relatively flat for lower mass galaxies 
($\log(M/M_\odot) = 9$) and steepen with increasing stellar mass.  In SDSS DR7, 
the spectroscopic fiber diameter is 3"; this corresponds to a minimum physical 
diameter of 1.31 $h^{-1}$kpc at a redshift $z < 0.03$.  For most of the dwarf 
galaxies in this study, this contains the majority of their angular size.  
Assuming the abundance gradients remain flat for dwarf galaxies, as suggested by 
the results of \cite{Belfiore17}, then our estimates of the gas-phase chemical 
abundances for our sample of dwarf galaxies are independent of the location of 
the spectral fiber on the galaxies' surfaces.

The selection criteria outlined in Section \ref{sec:SDSS_limits} limit our 
sample to only star-forming dwarf galaxies.  As a result, this is not a 
representative sample of the entire dwarf galaxy population.  We are only able 
to discuss the influence of the large-scale environment on star-forming dwarf 
galaxies in this study.  Unfortunately, it is impossible to estimate the 
chemical abundances of red dwarf galaxies with the direct $T_e$ method because 
the UV photons from young stars are needed to excite the interstellar gas.  
In addition, our temperature requirement eliminates extremely metal-poor 
galaxies.  To examine the influence of this bias on our results, we repeat the 
analysis with a maximum allowed temperature of $2.5\times 10^4$ and 
$3\times 10^4 \text{ K}$ \citep[the maximum allowed temperature in the work by]
[]{Douglass17a,Douglass17b}.  The resulting shifts in each abundance ratio are 
listed in Table \ref{tab:temp_test}.  There is no difference in the measured 
shifts seen with temperature cutoffs of $3\times 10^4$ and 
$2.5\times 10^4 \text{ K}$, but a significant number of low-metallicity galaxies 
are removed from the sample when we require temperatures no higher than 
$2\times 10^4 \text{ K}$.  While we still see shifts in the various abundance 
distributions, their statistical significance is reduced.

\floattable
\begin{deluxetable}{CCCC}
    \tablewidth{0pt}
    \tablecolumns{4}
    \tablecaption{Effect of maximum temperature on abundance shifts\label{tab:temp_test}}
    \tablehead{\colhead{Max $T$[\ion{O}{3}]} & \colhead{\OH} & \colhead{\NO} & \colhead{\NeO}}
    \startdata
        2\times 10^4 \text{ K}   & -0.01\pm 0.013\, (2\%) & 0.03\pm 0.016\, (7\%) & -0.02\pm 0.023\, (5\%)\\
        2.5\times 10^4 \text{ K} & -0.02\pm 0.013\, (5\%) & 0.04\pm 0.016\, (10\%) & -0.03\pm 0.023\, (7\%)\\
        3\times 10^4 \text{ K}   & -0.02\pm 0.013\, (5\%) & 0.04\pm 0.016\, (10\%) & -0.03\pm 0.023\, (7\%)\\
    \enddata
    \tablecomments{Magnitude of the average shifts between the void and wall 
    dwarf galaxies, with samples limited by the maximum temperature listed on 
    the left.  Positive shifts indicate that the wall values are greater than 
    the void values; negative shifts indicate that the void values are greater 
    than the wall values.  These shifts change very little with the different 
    maximum temperatures, so the bias toward higher metallicities introduced 
    with the maximum allowed temperature of $2\times 10^4 \text{ K}$ reduces the 
    significance of our results.}
\end{deluxetable}

%-------------------------------------------------------------------------------
\subsection{Comparison to previously published oxygen abundance estimates}

\begin{figure}
    \centering
    \includegraphics[width=0.5\textwidth]{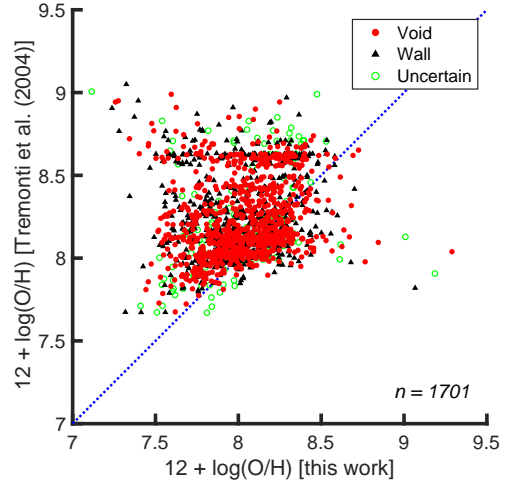}
    \caption{Oxygen abundance comparison between our calculated estimates with 
    the O$^+$ approximation and those made by \cite{Tremonti04}.  Error bars 
    have been omitted for clarity.  While the majority of our abundance 
    estimates agree reasonably well with the values already published, it is 
    clear that our estimates are often lower than the previously published 
    values.  It is well known that the strong-line methods \citep[like those 
    used by][]{Tremonti04} overestimate the oxygen abundance by as much as 0.3 
    dex \citep{Kennicutt03}.  Therefore, it is not surprising that the oxygen 
    abundances measured using the direct $T_e$ method are lower, particularly at 
    very low metallicities.}
    \label{fig:T04_comp}
\end{figure}

While no estimates of the nitrogen or N/O abundances have been made on a large 
selection of the SDSS galaxies, we can compare our oxygen abundance estimates to 
the metallicity values measured by \cite{Tremonti04}.  While we both use data 
from the MPA-JHU value-added catalog, \cite{Tremonti04} employs an empirical 
method to calculate the metallicity that is based on calibrated relationships 
between direct $T_e$ methods and strong-line ratios.  The results of this 
comparison are shown in Fig. \ref{fig:T04_comp}.  While the majority of our 
abundance estimates agree reasonably well with the values calculated by 
\cite{Tremonti04}, it is also clear that our estimates often predict abundances 
lower than those previously published.  This is especially true for the 
low-metallicity regime (\OH $< 7.6$).  Methods that are based on calibrations 
rarely use low-metallicity galaxies in their source for calibrating.  As a 
result, empirical methods will often overestimate the abundance values, 
especially in the low-metallicity regime.  \cite{Kennicutt03} show that 
strong-line methods (methods which make extreme use of the strong emission 
lines) can overestimate the metallicity abundances by as much as 0.3 dex.  
Therefore, we are not surprised at the apparent lack of correlation between our 
oxygen abundance estimates and those of \cite{Tremonti04}.

%-------------------------------------------------------------------------------
\subsection{N/O versus O/H} \label{sec:NO_OH}

\begin{figure}
    \centering
    \includegraphics[width=0.5\textwidth]{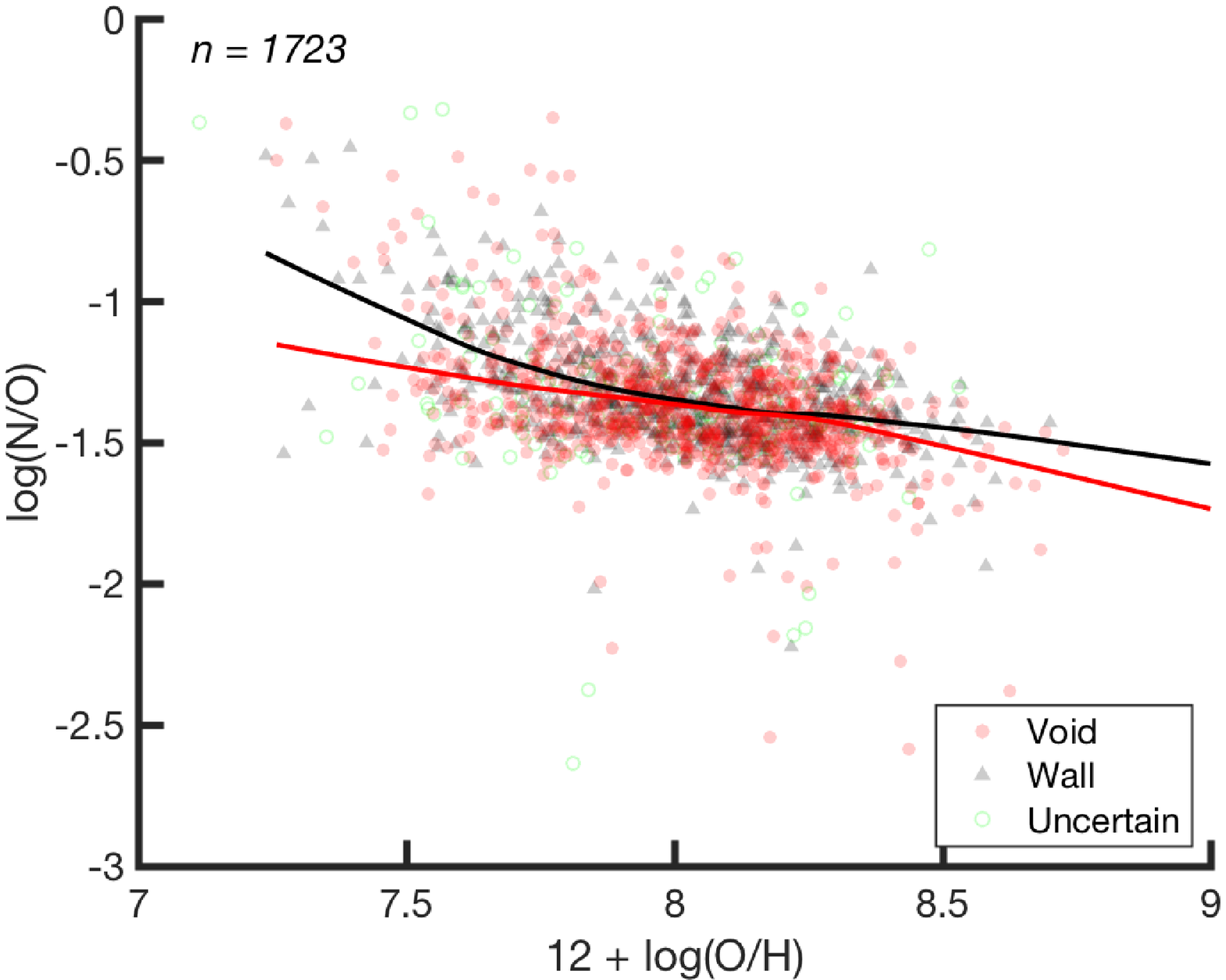}
    \caption{Oxygen abundance (\OH) versus the N/O ratio (\NO) for the dwarf 
    galaxies in this study; local linear regressions are also shown.  As seen in 
    \cite{Andrews13} and \cite{Douglass17b}, while the distribution exhibits a 
    roughly constant value of N/O as expected, the local linear regressions 
    reveal a slight negative correlation between the two abundance ratios.  
    There is no evidence of secondary nitrogen production in these samples of 
    star-forming dwarf galaxies, which would manifest as a positive relationship 
    between O/H and N/O at higher metallicities.}
    \label{fig:NOvOH}
\end{figure}

Studying how the N/O ratio depends on the metallicity (gas-phase oxygen 
abundance) probes the nucleosynthetic production of nitrogen in stars within the 
galaxies.  It is believed that nitrogen can be produced as both a primary and 
secondary element, depending on the initial metallicity of the stars.  If there 
are enough of the heavy elements available when the stars are created (oxygen, 
carbon, etc.), then the CNO cycle can commence much earlier in the star's 
lifetime, resulting in a higher production of nitrogen than if the star is 
originally created with very few heavy elements.  If this is the case, then we 
should see no relationship between the N/O ratio and the oxygen abundance below 
a certain metallicity value (the primary nitrogen production phase); above this 
threshold metallicity, the N/O value should increase linearly with the oxygen 
abundance (the secondary nitrogen production phase).

As we see in Fig. \ref{fig:NOvOH}, there is no evidence of a secondary nitrogen 
production phase in our sample of dwarf galaxies.  This is in contrast with the 
evidence of secondary nitrogen production seen in Fig. \ref{fig:M_NO} of Sec. 
\ref{sec:Mass}; the source of this discrepancy is unclear.  The relationship 
between metallicity and the N/O ratio instead shows a large scatter, where the 
N/O ratio is roughly independent of the metallicity.  This relationship between 
the N/O ratio and metallicity has been observed many times 
\citep[e.g.,][]{Garnett90, VilaCostas93, Thuan95, Izotov99, Henry00, Pilyugin02, 
Lee04, Pilyugin04, Nava06, vanZee06a, PerezMontero09, Amorin10, Berg12, 
SanchezAlmeida16}, and it has been interpreted to represent the production of 
primary nitrogen.  As listed in Table \ref{tab:stats}, we find that the void 
dwarf galaxies have a median N/O value of -1.38 and the wall dwarf galaxies have 
a median N/O value of -1.36.  These values fall within the previously reported 
N/O ratio averages for the plateau.

Also shown in Fig. \ref{fig:NOvOH} are local linear regressions of the void 
(red) and wall (black) dwarf galaxy samples.  The local linear regressions in 
this and the following figures are calculated at each point using a window 
containing 50\% of the nearest data.  These non-parametric approximations 
successfully represent the central trend of the data, but they exaggerate the 
sparse scatter in the tails.  Here they exhibit a slight negative correlation 
between the N/O ratio and the oxygen abundance.  This would indicate a constant 
value of nitrogen being synthesized in the galaxies, independent of the amount 
of oxygen being produced.  A negative correlation similar to this was also 
observed by \cite{Andrews13} and \cite{Douglass17b}.  In a footnote, 
\cite{Andrews13} note that they measure a slope of -0.21 for their stellar 
mass-binned galaxies with metallicities \OH $< 8.5$, while \cite{Douglass17b} 
find a slope of $-0.38\pm 0.078$ for dwarf galaxies.  Linear fits to the local 
linear regressions find slopes of $-0.254\pm 0.0029$ and $-0.36\pm 0.010$ for 
the void and wall dwarf galaxies, respectively.  We attribute these negative 
correlations to the significant scatter in the distributions and do not believe 
them to be physically relevant.

\begin{figure}
    \includegraphics[width=0.5\textwidth]{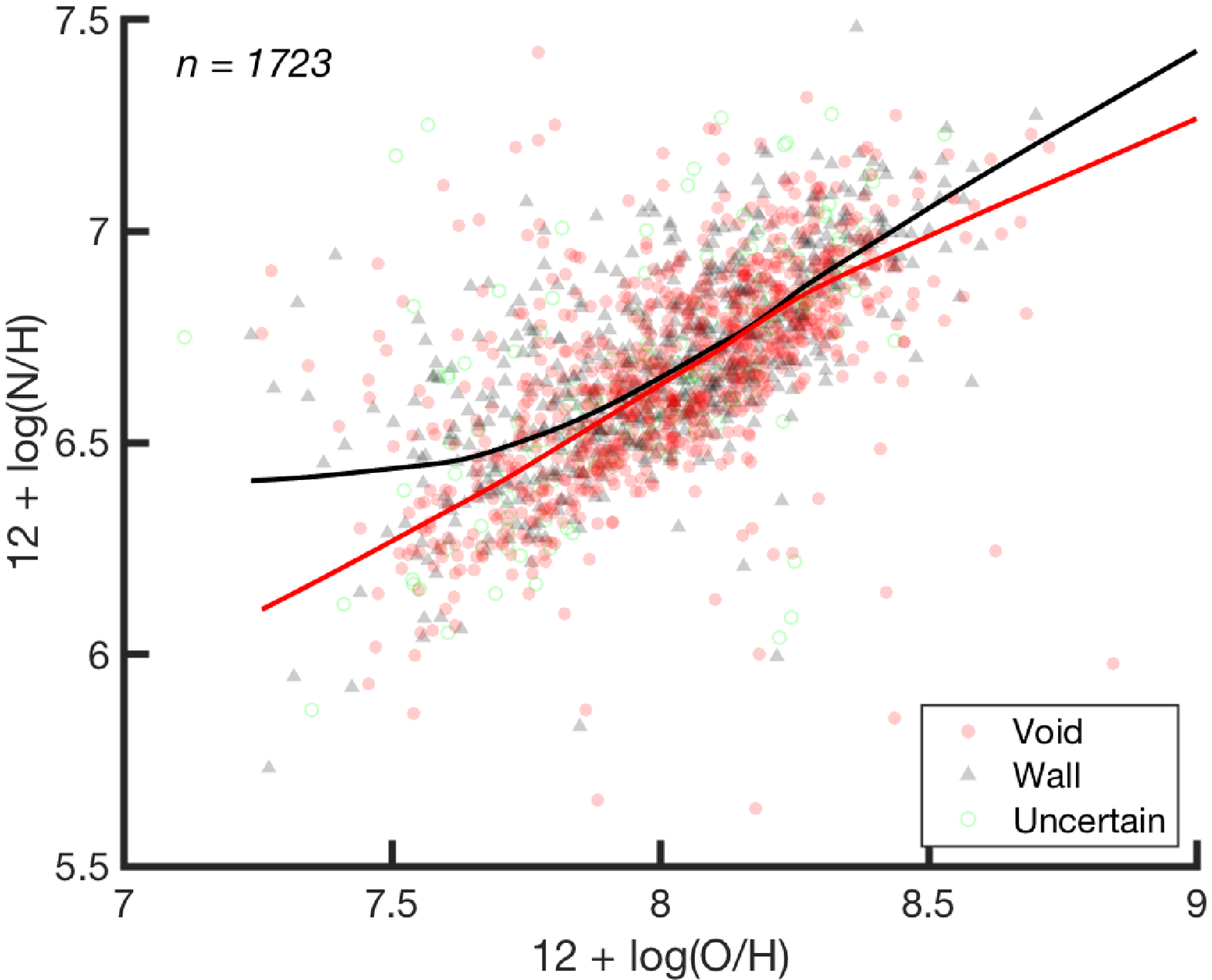}
    \caption{Oxygen abundance (\OH) versus nitrogen abundance (\NH) for the 
    dwarf galaxies in this study; local linear regressions are also shown.}
    \label{fig:OvN}
\end{figure}

Similar evidence of the nitrogen production phases can also be observed when 
looking at the relationship between the nitrogen and oxygen abundances relative 
to hydrogen.  Fig. \ref{fig:OvN} depicts a positive relationship between the 
nitrogen and oxygen abundances.  Linear fits to the central segments of the 
local linear regressions of the void and wall dwarf galaxy samples exhibit 
slopes of $0.746\pm 0.0029$ and $0.70\pm 0.034$, respectively, larger than the 
slope of $0.62\pm 0.078$ found by \cite{Douglass17b}.  These slopes indicate 
that nitrogen is produced at a slower rate than oxygen in dwarf galaxies.  A 
nitrogen plateau in Fig. \ref{fig:NOvOH} would manifest itself as a relationship 
between the N/H and O/H ratios (shown in Fig. \ref{fig:OvN}) with a linear slope 
of 1.  Secondary nitrogen production, a positive relationship in Fig. 
\ref{fig:NOvOH}, would correspond to a relationship between N/H and O/H in Fig. 
\ref{fig:OvN} with a slope greater than 1.

The slopes of the local linear regressions in both Figs. \ref{fig:NOvOH} and 
\ref{fig:OvN} show that there is no significant difference in the abundance 
ratio relationships between void dwarf galaxies and dwarf galaxies in denser 
regions.  This supports the results found by \cite{Douglass17b} and 
\cite{Vincenzo18}.  The large-scale environment does not appear to have an 
influence in the nucleosynthesis of nitrogen in dwarf galaxies.

%-------------------------------------------------------------------------------
\subsection{Ne/O versus O/H} \label{sec:NeO_OH}

\begin{figure}
    \centering
    \includegraphics[width=0.5\textwidth]{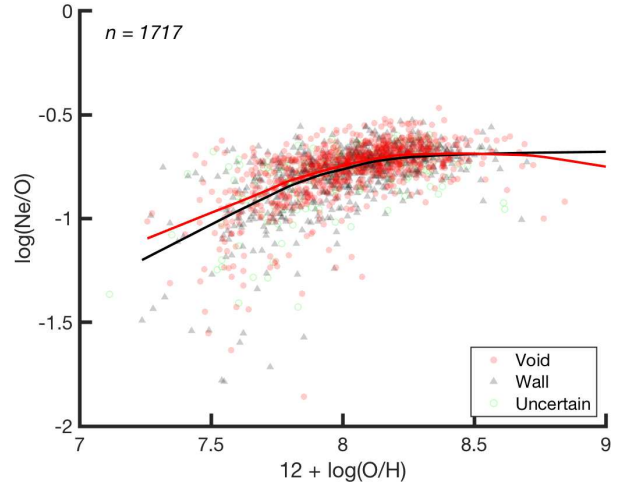}
    \caption{Oxygen abundance (\OH) versus the Ne/O ratio (\NeO) for the dwarf 
    galaxies in this study; local linear regressions are also shown.}
    \label{fig:NeOvOH}
\end{figure}

To serve as a diagnostic for the abundance estimates, we also look at the 
relationship between the metallicity and the Ne/O ratio.  Because both neon and 
oxygen are synthesized in $\alpha$ processes, their relative abundance should 
not depend on a galaxy's metallicity.  As Fig. \ref{fig:NeOvOH} shows, there is 
no significant relationship between the oxygen abundance and the Ne/O ratio.  
This matches the results of \cite{Kobulnicky96,Izotov99,Lee04,Izotov06,
vanZee06a} and \cite{PerezMontero07}.  We note an increase in the scatter of 
this relationship with decreasing metallicity, which the local linear 
regressions shown in Fig. \ref{fig:NeOvOH} emphasize.  Since both the 
[\ion{O}{3}] $\lambda$4363 and the [\ion{Ne}{3}] $\lambda$3869 emission lines 
are relatively weak, the quality of their detection decreases with the 
metallicity and therefore increases the scatter in this relationship.  The local 
linear regressions of the void and wall dwarf galaxies in Fig. \ref{fig:NeOvOH} 
do not show a difference in the abundance ratio relationship between void dwarf 
galaxies and dwarf galaxies in denser regions.  As expected from stellar 
nucleosynthesis, the void environment does not appear to influence the relative 
abundances of $\alpha$-process particles in dwarf galaxies.

%-------------------------------------------------------------------------------
\subsection{Stellar mass--Abundance relations}\label{sec:Mass}

\begin{figure*}
    \centering
    \includegraphics[width=0.49\textwidth]{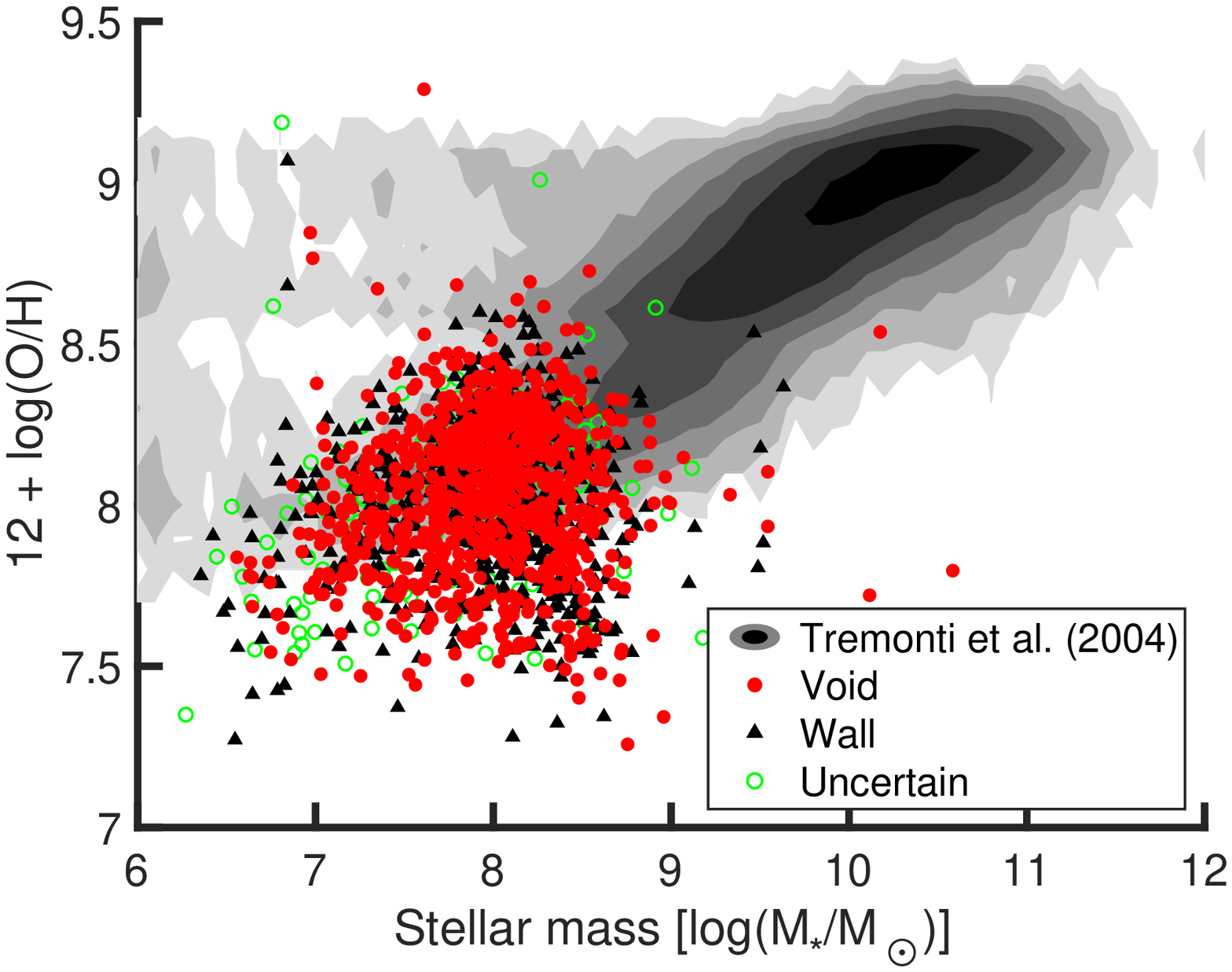}
    \includegraphics[width=0.49\textwidth]{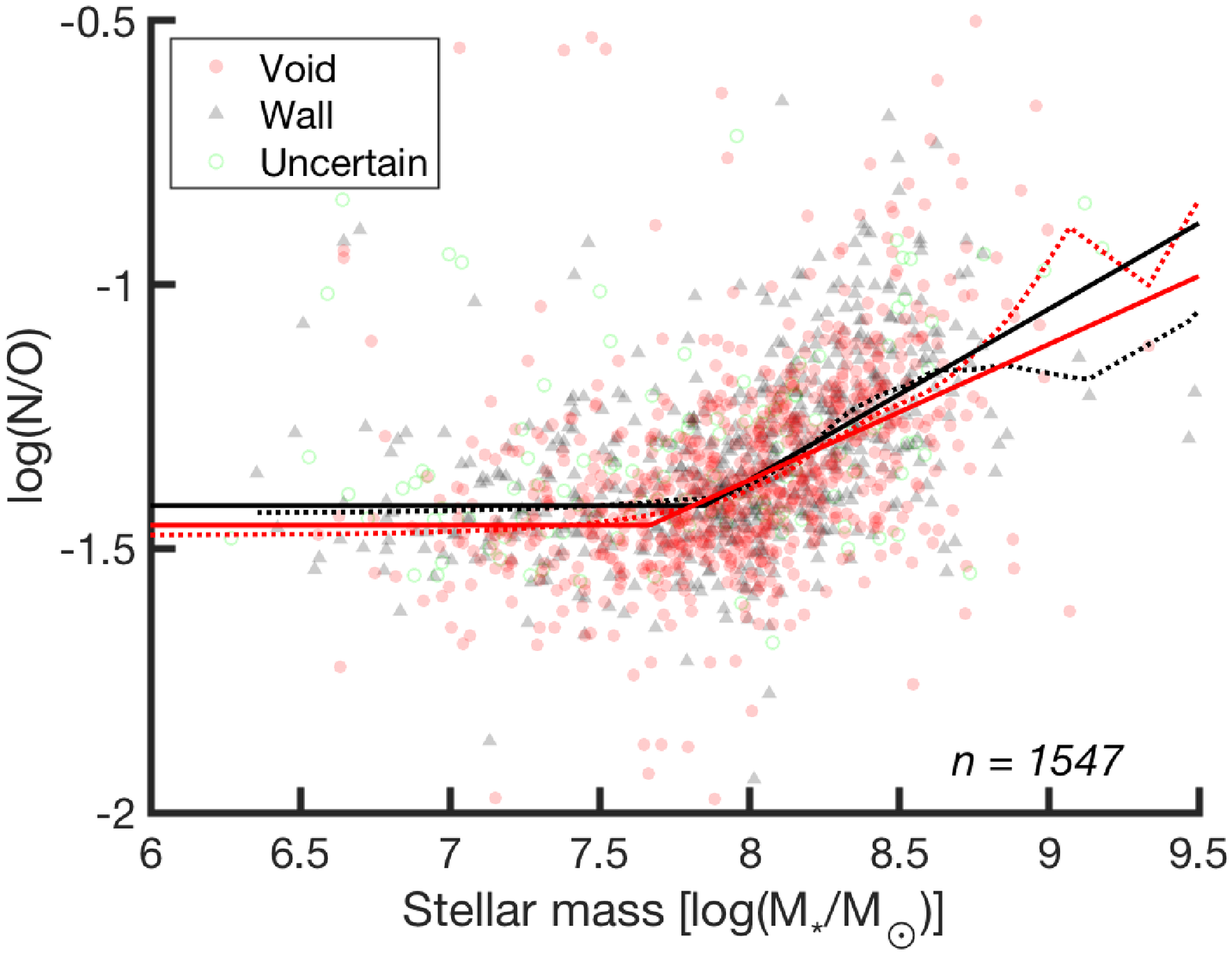}
    \caption{Stellar mass versus oxygen abundance (left panel) and N/O ratio 
    (right panel).  We include the metallicity results of \cite{Tremonti04} on 
    the left to place our dwarf galaxy abundance results in context.  On the 
    right, we show local linear regressions (dashed lines) and piecewise linear 
    functions fit to the regressions (solid lines) to the two dwarf galaxy 
    populations.  Most of our dwarf galaxies follow the same mass-metallicity 
    relationship seen in \cite{Tremonti04}.  There is no clear difference 
    between void and wall dwarf galaxies in the mass-metallicity relation.  In 
    the right-hand panel, though, we see that the turn-off for the N/O plateau 
    occurs at different masses for the two large-scale environments.}
    \label{fig:M_NO}
\end{figure*}

Expanding on the mass-metallicity relation investigated in \cite{Douglass17a}, 
we look at the correlation between stellar mass and the oxygen abundance in our 
now substantially larger sample of dwarf galaxies.  The mass-metallicity 
relation for our dwarf galaxies can be seen in the left panel of Fig. 
\ref{fig:M_NO}.  We also include those galaxies from the MPA-JHU catalog with 
metallicity estimates from \cite{Tremonti04} to place our sample in context.  
The majority of our dwarf galaxies follow the expected mass-metallicity trend.  
There is no discernible influence from the large-scale environment on the 
mass-metallicity relation for dwarf galaxies.

We also look at the N/O ratio as a function of stellar mass (Fig. 
\ref{fig:M_NO}, right panel).  Similar to the O/H--N/O relation studied in Sec. 
\ref{sec:NO_OH}, the N/O ratio is predicted to be constant below some critical 
mass; above this, the N/O ratio should increase linearly with the stellar mass.  
Unlike the relation seen in Fig. \ref{fig:NOvOH}, we see both the N/O plateau 
and the positive correlation on the right in Fig. \ref{fig:M_NO}.  To better 
investigate the influence of the void environment on the relationship between 
the stellar mass and the N/O ratio, we include local linear regressions of the 
two dwarf galaxy populations in Fig. \ref{fig:M_NO}.  We observe a difference in 
this relation as a function of the large-scale environment: piecewise linear 
functions fit to the local linear regressions reveal that the critical mass for 
void galaxies is around $\log(M_*/M_\odot) \sim 7.7$, while the galaxies in 
denser regions exhibit a critical mass of $\sim$7.8.  This difference suggests 
that void galaxies begin to synthesize secondary nitrogen at lower stellar 
masses than galaxies in more dense regions, consistent with the statistically 
insignificant shift seen in Fig. \ref{fig:met1sig} where void dwarf galaxies may 
have slightly higher oxygen abundances than dwarf galaxies in more dense 
environments.

%-------------------------------------------------------------------------------
\subsection{\ion{H}{1} mass--Abundance relations}

\begin{figure*}
    \centering
    \includegraphics[width=0.49\textwidth]{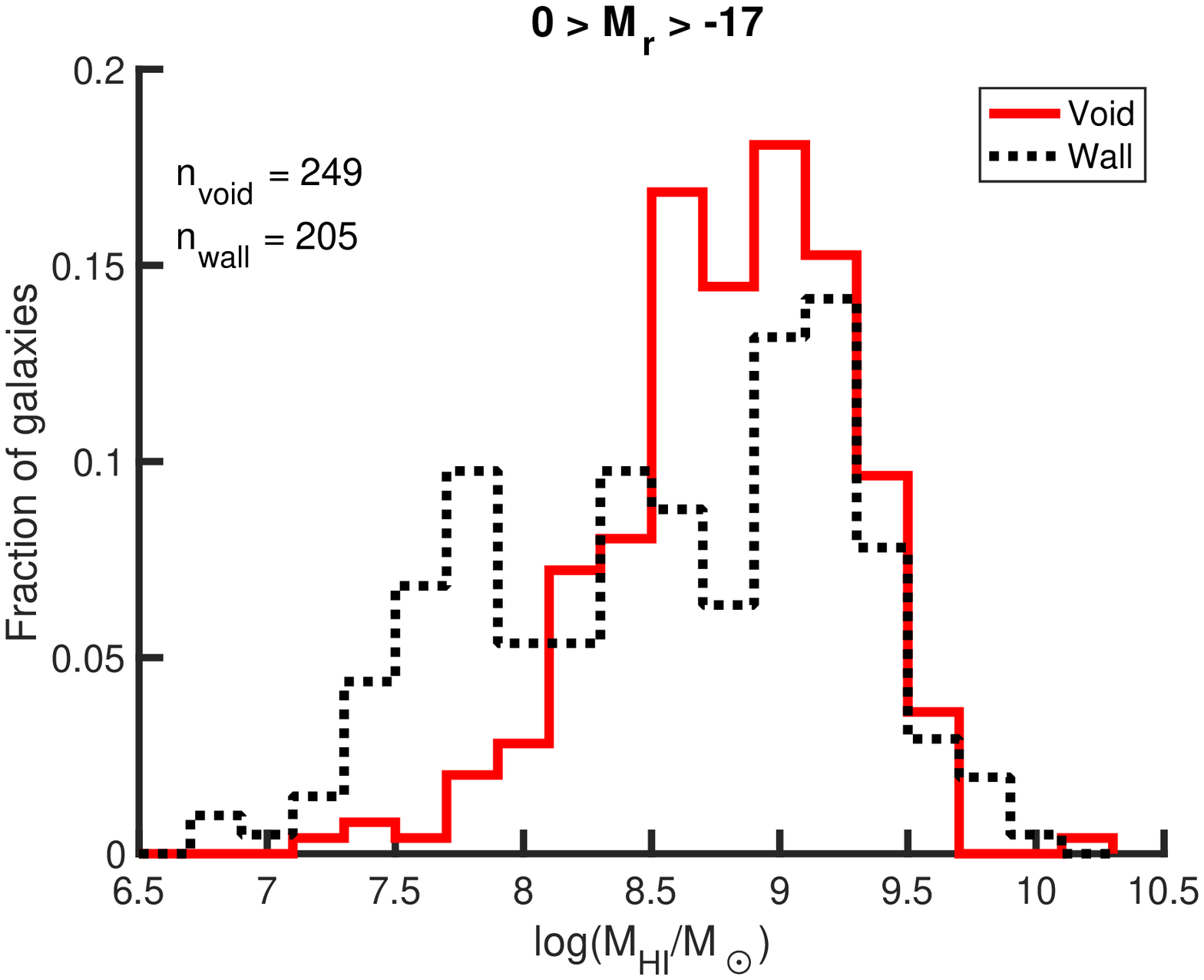}
    \includegraphics[width=0.49\textwidth]{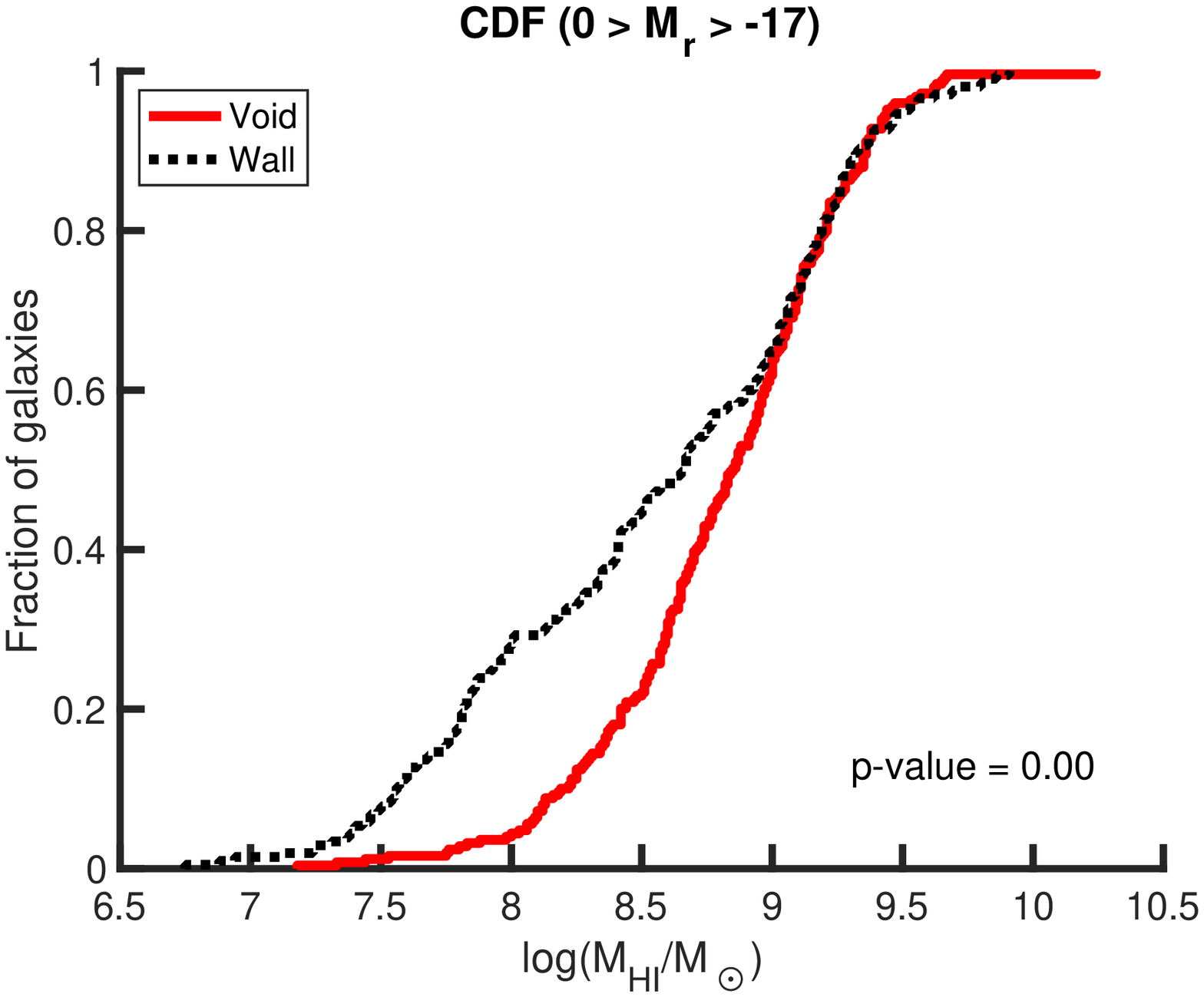}
    \caption{Distribution of \ion{H}{1} mass in the sample of dwarf galaxies, 
    separated by their large-scale environment.  There is an obvious shift 
    toward higher \ion{H}{1} masses in the void dwarf galaxies.}
    \label{fig:HI_hist}
\end{figure*}

In addition to looking at the stellar mass, we also investigate the relationship 
between the amount of neutral hydrogen and the gas-phase chemical abundances in 
our sample of star-forming dwarf galaxies.  As shown in \cite{Moorman14}, void 
galaxies typically have lower \ion{H}{1} masses than galaxies in more dense 
regions, consistent with the overall shift of the luminosity or stellar mass 
function to lower luminosity or mass in voids.  In the fixed range of luminosity 
of our dwarf galaxy sample, as Fig. \ref{fig:HI_hist} shows, our sample of void 
dwarf galaxies have higher \ion{H}{1} masses than the dwarf galaxies in denser 
regions.  The gas in the void environment is cooler than that found in denser 
regions (due to less events like shock heating, etc.), which permits void 
galaxies to have higher \ion{H}{1} masses for a given stellar mass.  Therefore, 
because we are fixing the stellar mass in our sample of galaxies (by only 
studying dwarf galaxies), we should find that the void dwarf galaxies have 
higher \ion{H}{1} masses than for those dwarf galaxies in denser environments.

\begin{figure*}
    \centering
    \includegraphics[width=0.49\textwidth]{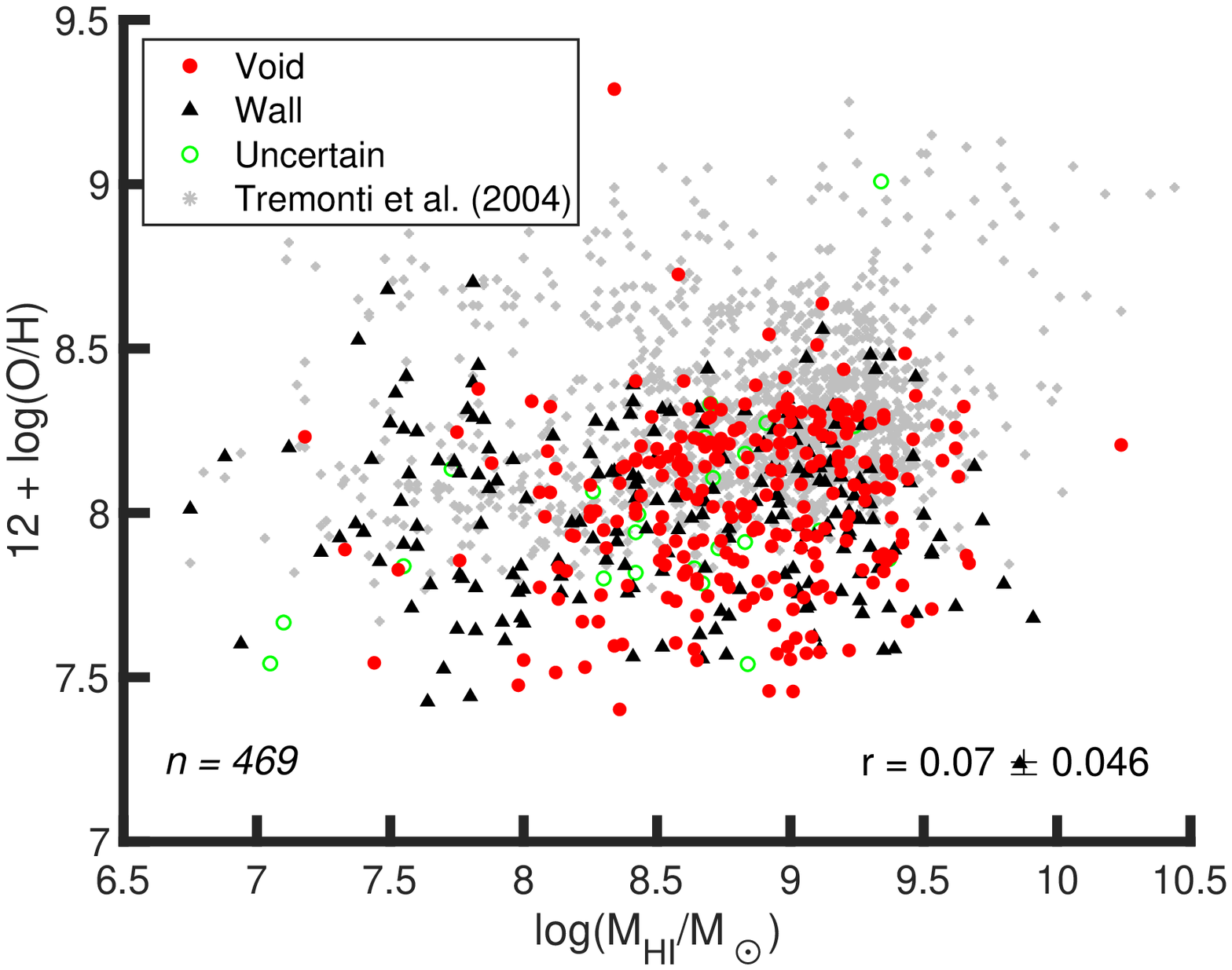}
    \includegraphics[width=0.49\textwidth]{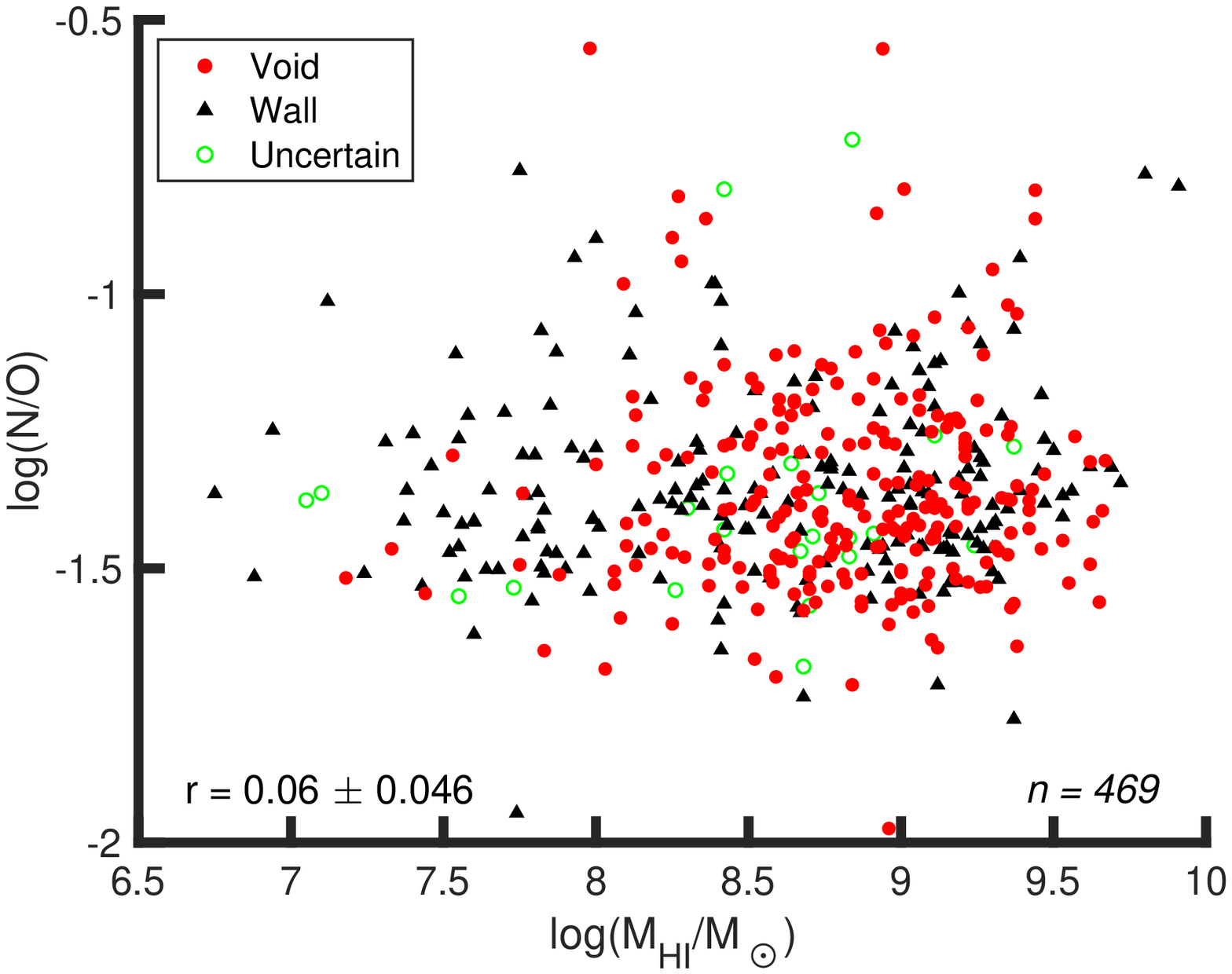}
    \includegraphics[width=0.49\textwidth]{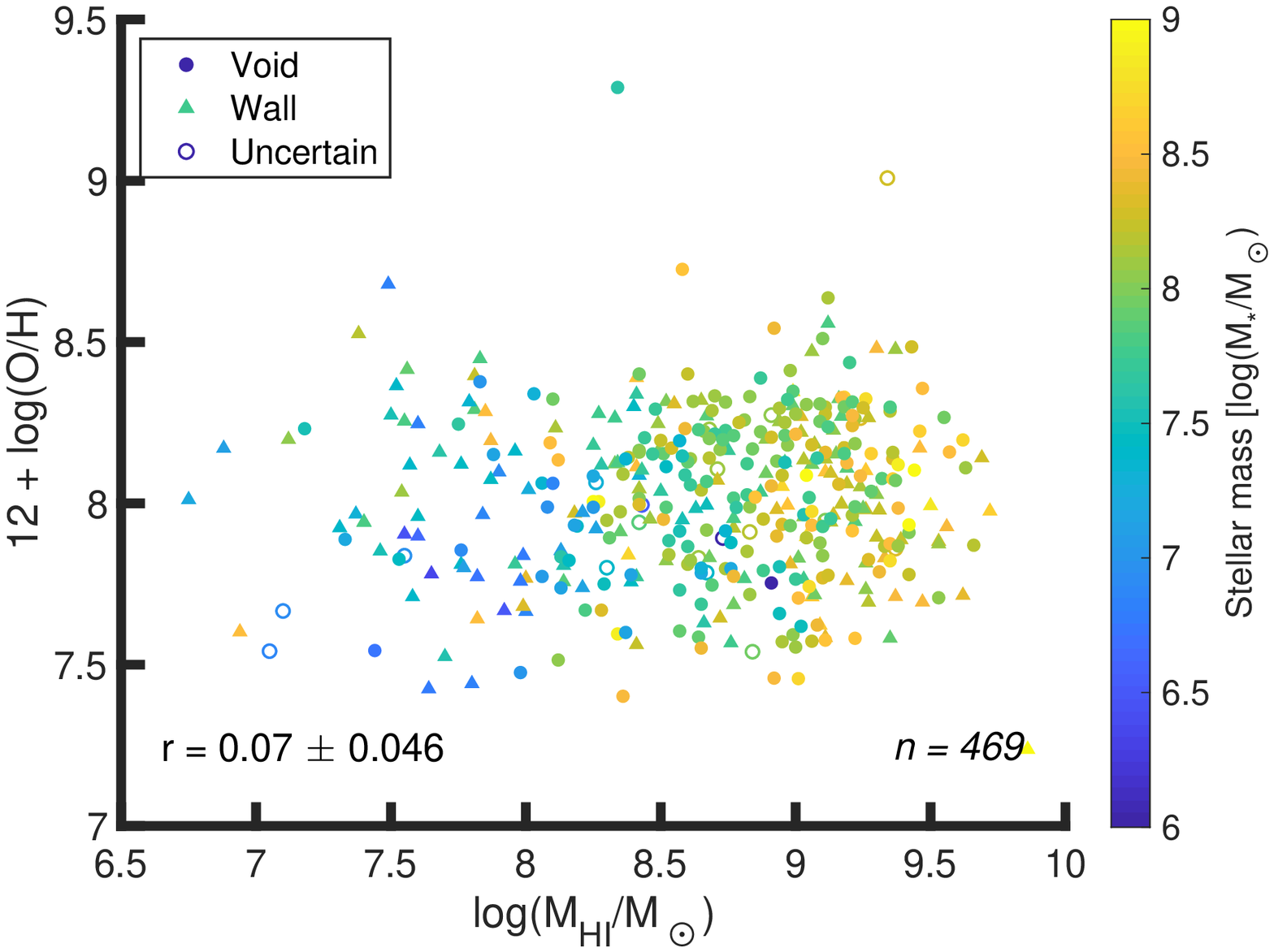}
    \includegraphics[width=0.49\textwidth]{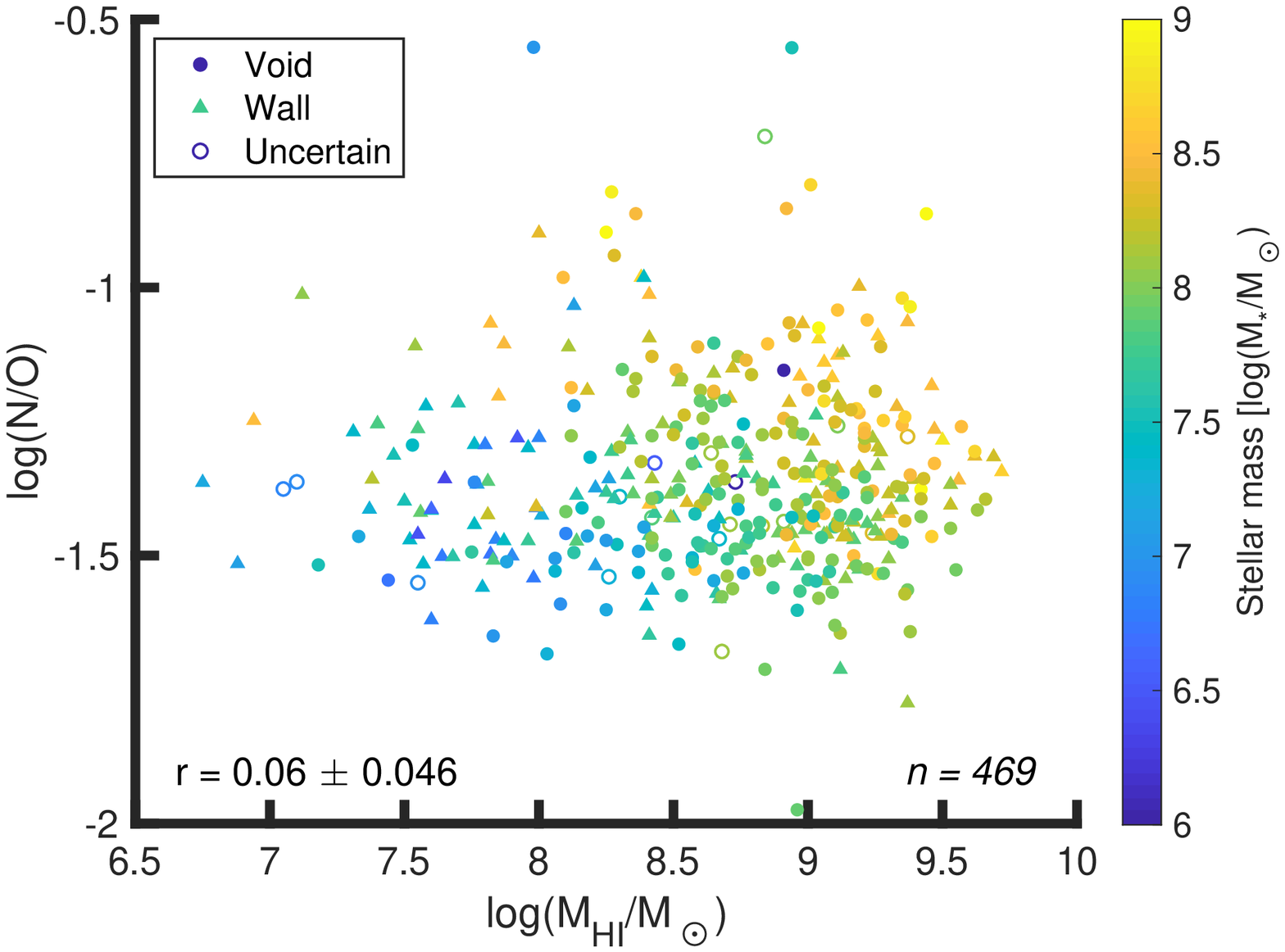}
    \caption{\ion{H}{1} mass versus metallicity (left) and N/O ratio (right) for 
    star-forming dwarf galaxies.  The color scheme of the top row emphasizes the 
    large-scale environment of the star-forming dwarf galaxies, while the bottom 
    row investigates the relationship between stellar mass, \ion{H}{1} mass, and 
    chemical abundance.  Error bars have been omitted for clarity.  To place our 
    oxygen abundance results in context, we show (gray stars) the dwarf galaxies 
    in SDSS DR7 with metallicity estimates from \cite{Tremonti04}.}
    \label{fig:HI}
\end{figure*}

The \ion{H}{1} mass--metallicity and \ion{H}{1} mass--N/O relations can be seen 
in Fig. \ref{fig:HI}, extending the results of \cite{Bothwell13} down to 
$\log(M_*/M_\odot) \approx 6$.  Unlike the correlation between the stellar mass 
and the chemical abundances, there is very little correlation between the 
abundances and the \ion{H}{1} mass.  There is no significant relationship 
between the \ion{H}{1} mass and the metallicity of a galaxy --- the 
correlation coefficient for the dwarf galaxies shown in the upper left panel of 
Fig. \ref{fig:HI} is only $0.07\pm 0.046$.  This is not surprising, as our 
sample of star-forming dwarf galaxies probes the low-metallicity range of the MZ 
relation where there is no strong relationship between the metallicity and 
stellar mass.  We also see no significant relationship between a dwarf galaxy's 
\ion{H}{1} mass and its N/O ratio; the correlation coefficient for the galaxies 
shown in the upper right panel of Fig. \ref{fig:HI} is only $0.06\pm 0.046$.

Due to the time delay in the production of nitrogen relative to oxygen, we 
expect the N/O ratio to decrease with increasing \ion{H}{1} mass, since more 
evolved galaxies have higher N/O ratios and will have used up most of their 
neutral hydrogen.  \cite{Bothwell13} demonstrates the existence of this 
relationship at a fixed stellar mass.  The first row of Fig. \ref{fig:HI} shows 
little, if any, relationship between the \ion{H}{1} mass and the chemical 
abundances.  However, the bottom row of Fig. \ref{fig:HI} shows that there is an 
inverse relationship between the chemical abundance and the \ion{H}{1} mass for 
fixed stellar mass (indicated by the colors of the points).

%-------------------------------------------------------------------------------
\subsection{Color--Abundance relations}

\begin{figure*}
    \centering
    \includegraphics[width=0.49\textwidth]{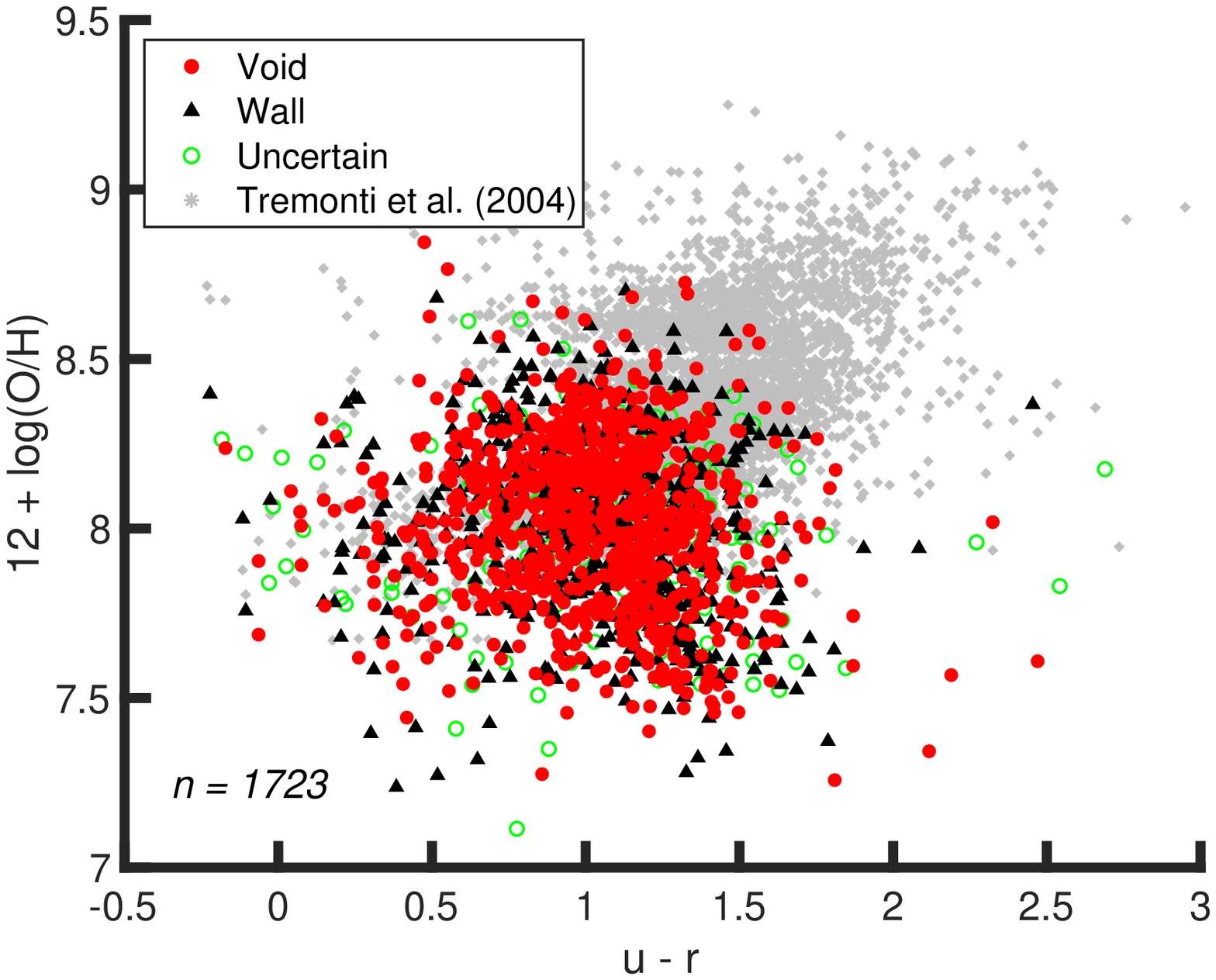}
    \includegraphics[width=0.49\textwidth]{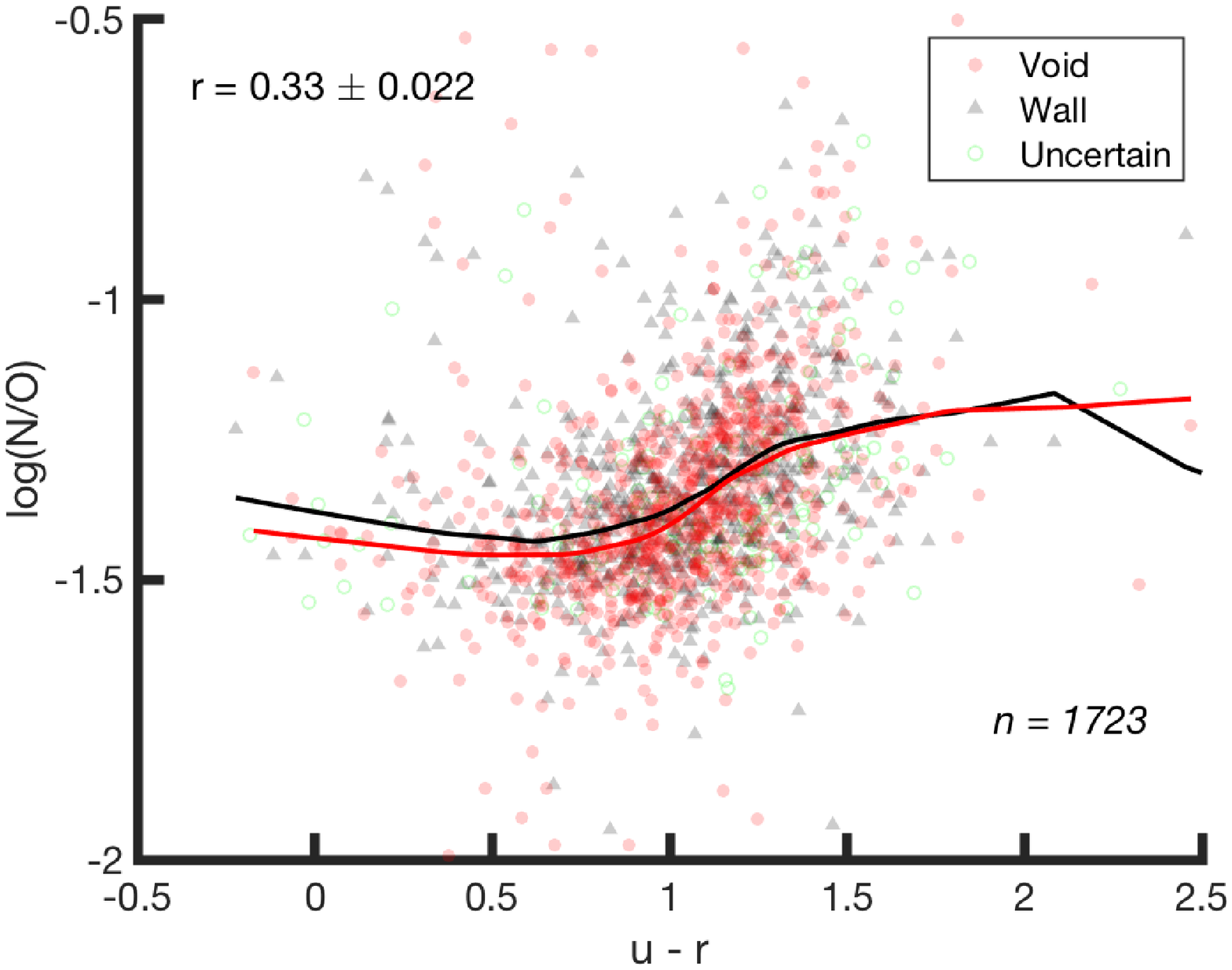}
    \caption{Color ($u-r$) versus the gas-phase oxygen abundance (left) and the 
    N/O ratio (right) for star-forming dwarf galaxies.  Error bars on individual 
    points have been omitted for clarity.  For reference, the dwarf galaxies in 
    SDSS DR7 that have estimated metallicities from \cite{Tremonti04} are shown 
    in the left panel in gray.  To discern any environmental trends in the 
    results, the local linear regressions of the two dwarf galaxy populations 
    are shown on the right.}
    \label{fig:ur}
\end{figure*}

\begin{figure*}
    \centering
    \includegraphics[width=0.49\textwidth]{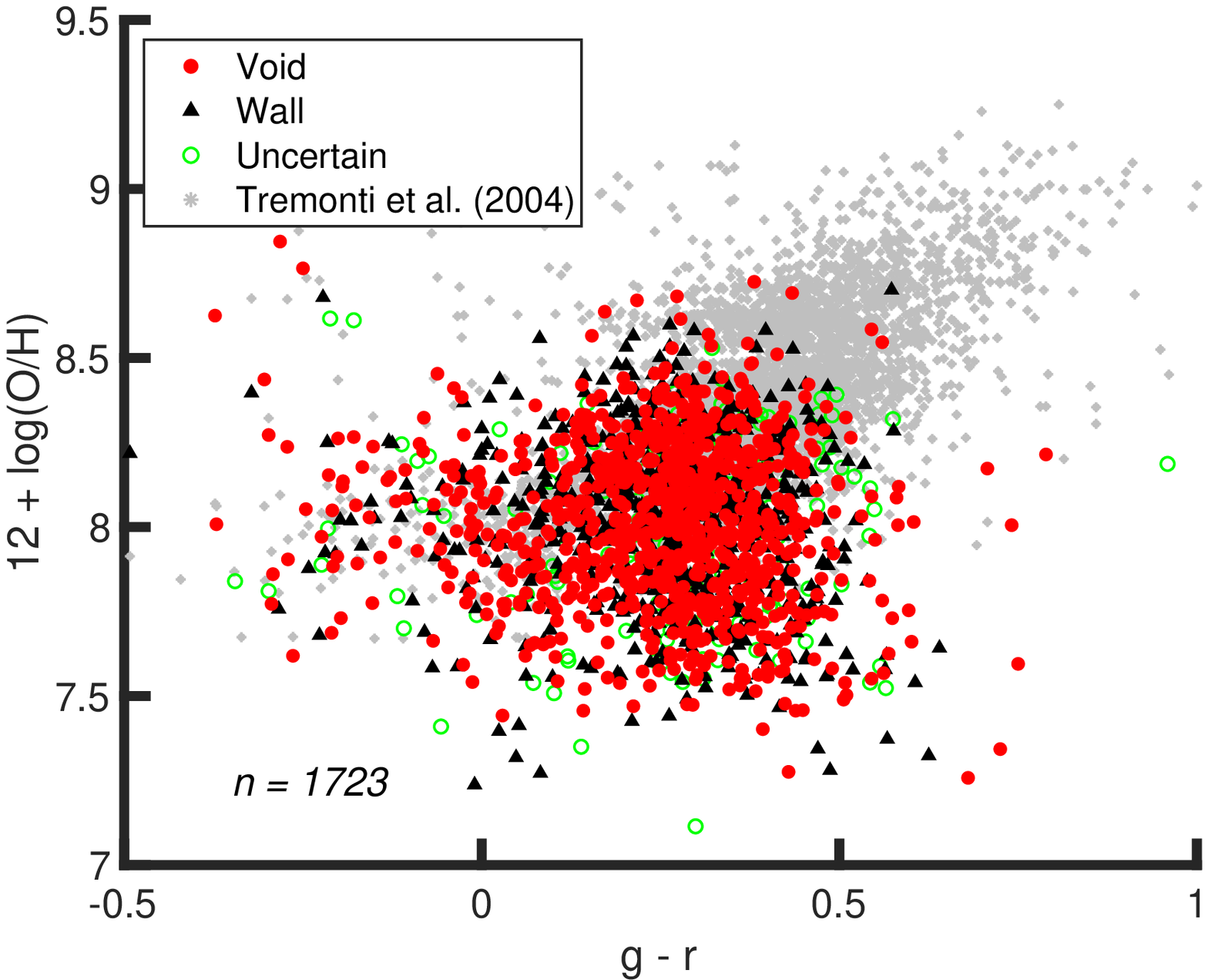}
    \includegraphics[width=0.49\textwidth]{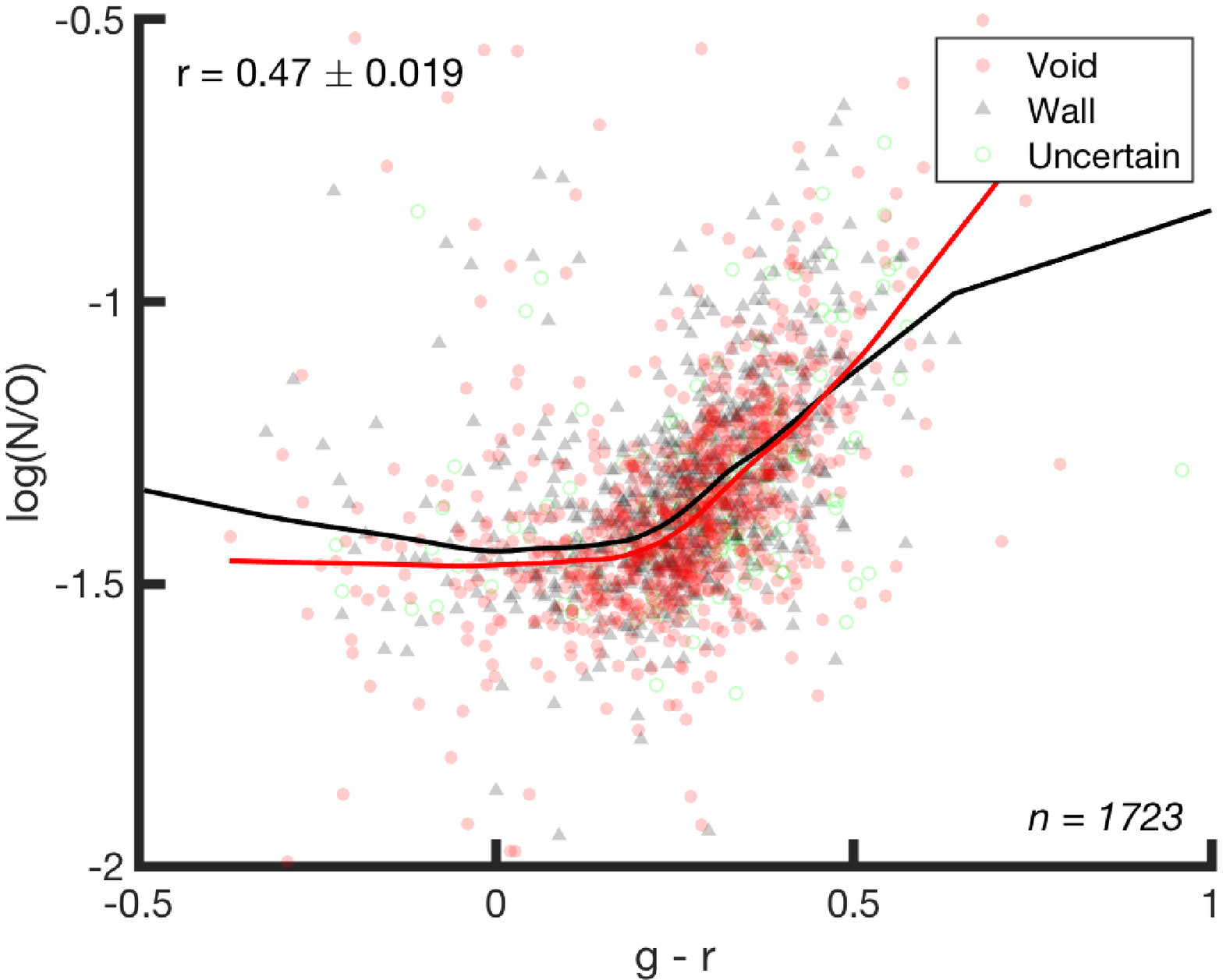}
    \caption{Color ($g-r$) versus the gas-phase oxygen abundance (left) and the 
    N/O ratio (right) for star-forming dwarf galaxies.  Error bars on individual 
    points have been omitted for clarity.  For reference, the galaxies in SDSS 
    DR7 that have estimated metallicities from \cite{Tremonti04} are shown in 
    the left panel in gray.  To discern any environmental trends in the results, 
    the local linear regressions of the two dwarf galaxy populations are shown 
    on the right.}
    \label{fig:gr}
\end{figure*}

The gas-phase chemical abundance is expected to have a positive correlation with 
a galaxy's color.  Older galaxies have had more time to convert their gas into 
heavier elements through star formation, increasing their metallicities.  The 
color--metallicity and color--N/O relations for our sample of star-forming dwarf 
galaxies can be seen in Figures \ref{fig:ur} and \ref{fig:gr}.  As we see, bluer 
galaxies have lower O/H and N/O ratios when we look at both the $u-r$ and $g-r$ 
colors.

The presence of a relationship between the N/O ratio and the color of a galaxy 
can indicate a time delay between the release of nitrogen and oxygen 
\citep{vanZee06a,Berg12}.  If higher-mass stars are the main source of oxygen, 
then the oxygen will be released on a shorter time scale than nitrogen for a 
given star formation episode (since higher-mass stars turn off the main sequence 
earlier than the intermediate-mass stars that synthesize nitrogen).  Therefore, 
the amount of nitrogen relative to oxygen should increase as the hotter, more 
massive stars burn out and the galaxy becomes redder.  This trend can be seen in 
the star-forming dwarf galaxies on the right of Figures \ref{fig:ur} and 
\ref{fig:gr}, matching the trends seen in \cite{Douglass17b,vanZee06a,Berg12}.

There does not appear to be any influence from the large-scale environment on 
the relationship between the color and chemical abundances for star-forming 
dwarf galaxies.  There is no obvious difference between the void and wall dwarf 
galaxies in the left-hand panels of Figures \ref{fig:ur} and \ref{fig:gr}, when 
we concentrate on the oxygen abundance as a function of color.  To help discern 
any influence from the environment on the N/O ratio as a function of color, we 
show the local linear regressions on the right-hand plots of Figures 
\ref{fig:ur} and \ref{fig:gr}.  The shift toward higher N/O ratios seen in the 
wall bins in both figures is the same shift identified in the histograms in Fig. 
\ref{fig:NOratio}.  Any variation in the color--abundance relationship between 
the void and wall populations except a vertical shift would be evidence of the 
large-scale environment influencing the relationship.  The scatter of the redder 
galaxies prohibits us from being able to make any assertions about differences 
in the slopes of these local linear regressions.

%-------------------------------------------------------------------------------
\subsection{(s)SFR--Abundance relations}

\begin{figure*}
    \centering
    \includegraphics[width=0.49\textwidth]{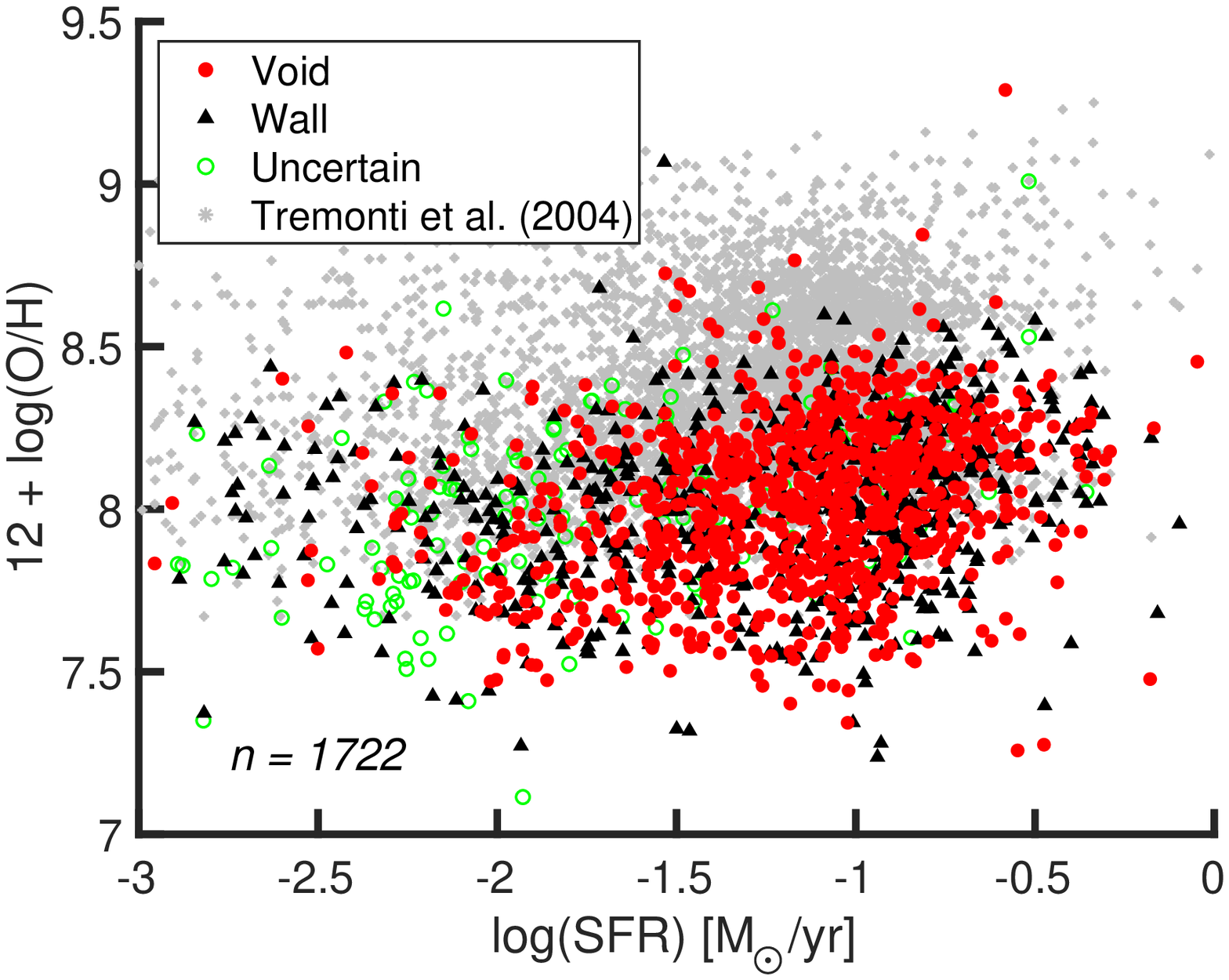}
    \includegraphics[width=0.49\textwidth]{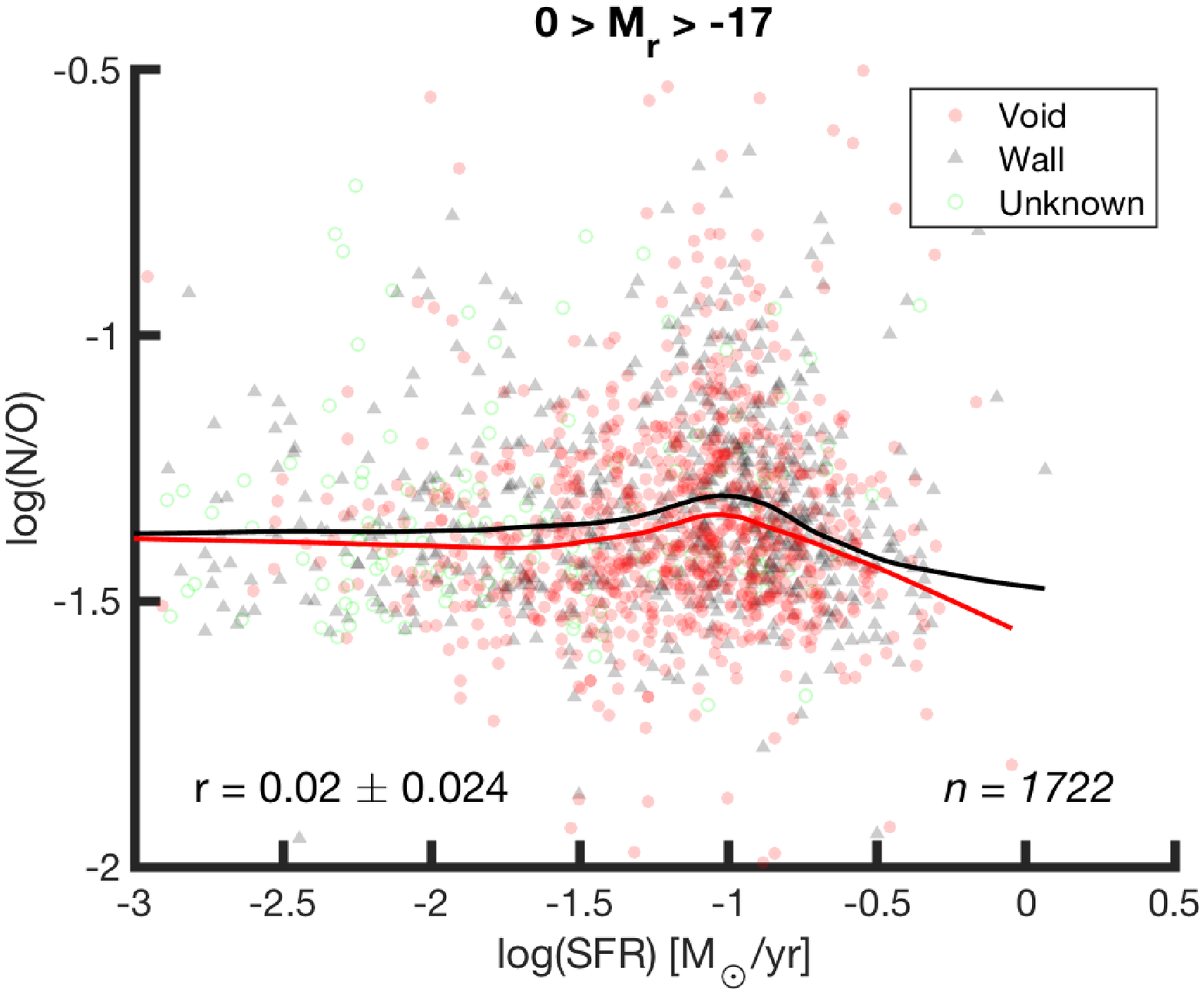}
    \caption{SFR versus metallicity (left) and the N/O ratio (right) for 
    star-forming dwarf galaxies.  Error bars for individual galaxies have been 
    omitted for clarity.  To place our oxygen abundance results in context, we 
    show (gray stars) the dwarf galaxies in SDSS DR7 with metallicity estimates 
    from \cite{Tremonti04}.  To discern any environmental effects on the 
    relation between SFR and the N/O ratio, the local linear regressions of the 
    two dwarf galaxy populations are shown on the right.}
    \label{fig:SFR}
\end{figure*}

\begin{figure*}
    \centering
    \includegraphics[width=0.49\textwidth]{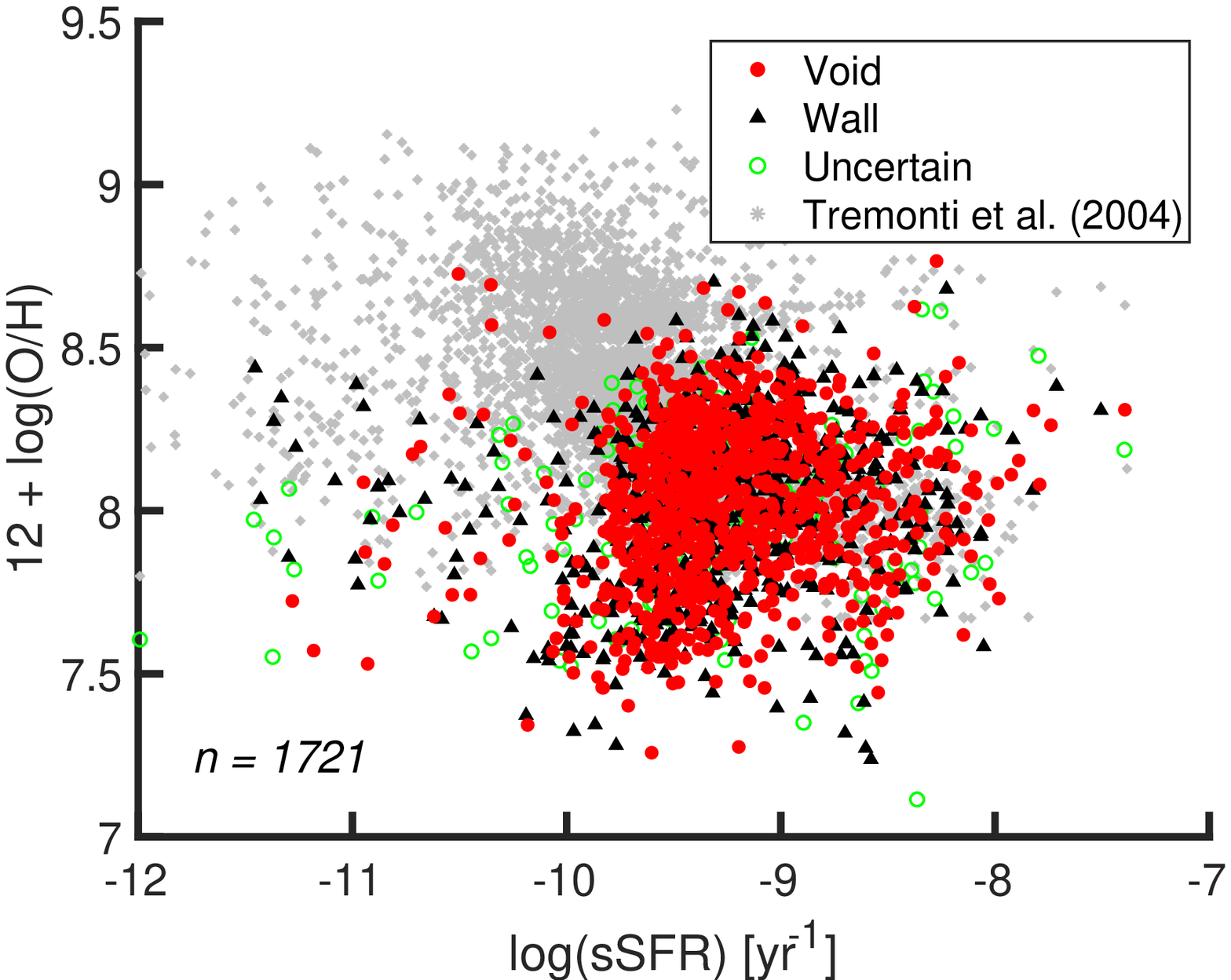}
    \includegraphics[width=0.49\textwidth]{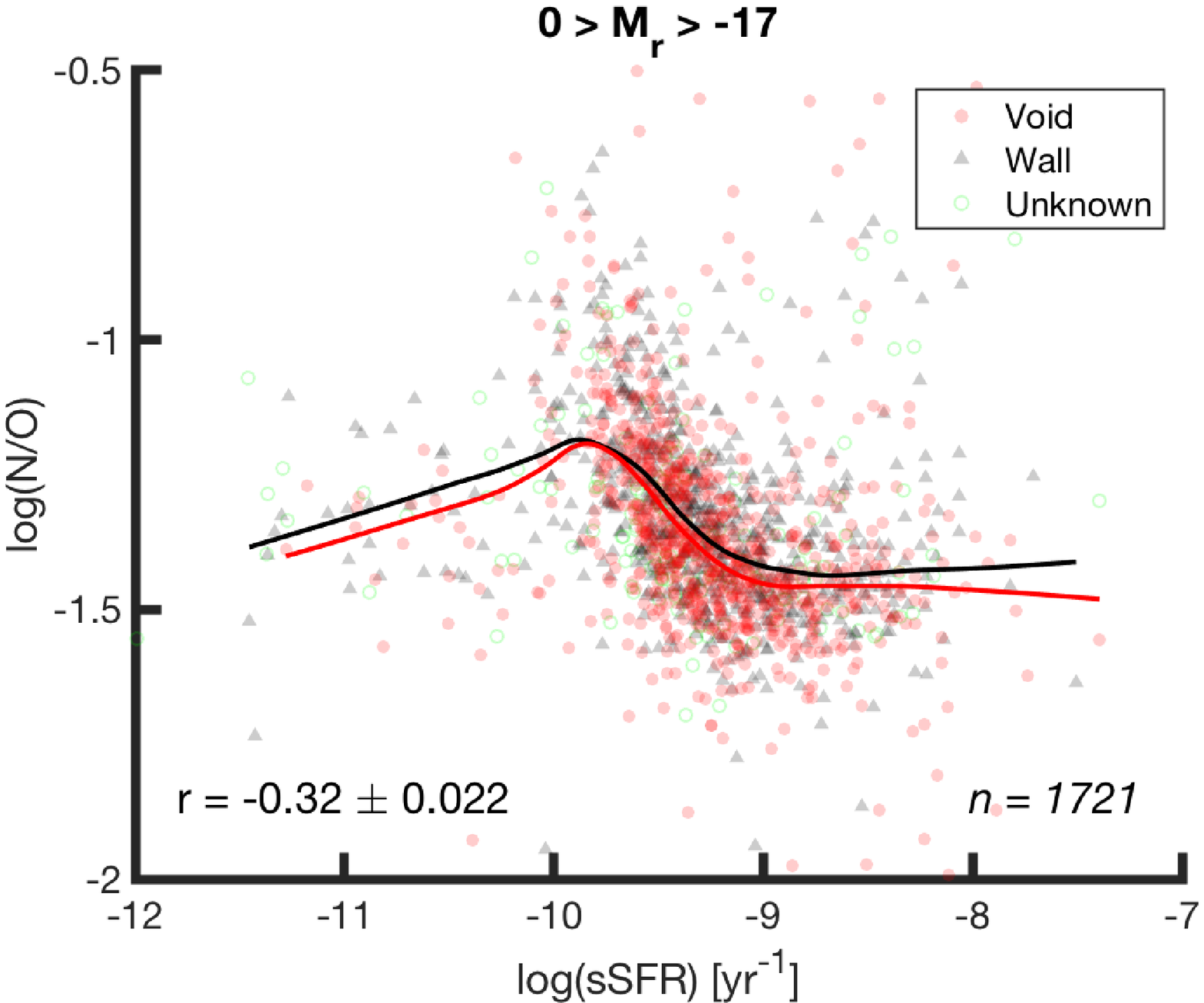}
    \caption{Specific star formation rate (sSFR) versus metallicity (left) and 
    the N/O ratio (right) for star-forming dwarf galaxies.  Error bars on 
    individual galaxies have been omitted for clarity.  To place our oxygen 
    abundance results in context, we show (gray stars) the dwarf galaxies in 
    SDSS DR7 with metallicity estimates from \cite{Tremonti04}.  To discern any 
    environmental effects on the relation between sSFR and the N/O ratio, the 
    local linear regressions of the two dwarf galaxy populations are shown on 
    the right.}
    \label{fig:sSFR}
\end{figure*}

There is thought to be a fundamental relationship between the stellar mass, star 
formation rate (SFR), and metallicity of a galaxy \citep{LaraLopez10,Mannucci10,
Andrews13}; the metallicity of a galaxy should increase with stellar mass and 
decrease as a function of the SFR.  \cite{Henry13} observe an inverse 
relationship between the metallicity and SFR of low-mass galaxies.  However, as 
was seen in \cite{Douglass17a}, Fig. \ref{fig:SFR} shows very little correlation 
between the SFR and metallicity or N/O ratio of the star-forming dwarf galaxies.  
When we separate the dwarf galaxies by their large-scale environment, we see no 
difference in the correlation coefficients between the two environments; there 
is no discernible influence on the relationship between the SFR and the chemical 
abundances by the large-scale environment.

We also inspect the relationship between the sSFR and the gas-phase chemical 
abundances in star-forming dwarf galaxies.  As shown in Fig. \ref{fig:sSFR}, 
there is a stronger correlation between the sSFR of a dwarf galaxy and its 
metallicity and N/O ratio.  The left-hand panel of Fig. \ref{fig:sSFR} shows the 
relationship between the gas-phase oxygen abundance and the sSFR for 
star-forming dwarf galaxies; to place our results in context, we also include 
(gray stars) those dwarf galaxies in SDSS DR7 for which \cite{Tremonti04} was 
able to estimate metallicities.  In the metallicity regime we are able to probe 
($12 + \log (\text{O}/\text{H}) \leq 8.5$), the star-forming dwarf galaxies 
display relatively little relationship between their sSFR and metallicity.

As we see on the right in Fig. \ref{fig:sSFR}, though, there is a strong 
anti-correlation between the sSFR and N/O ratio for the star-forming dwarf 
galaxies.  Galaxies with higher sSFRs may be producing more massive stars than 
galaxies with lower sSFRs.  If oxygen is produced in more massive stars than 
those which produce nitrogen, then the galaxies with higher sSFRs will produce 
more oxygen earlier than those galaxies with lower sSFRs.  Increasing the 
gas-phase oxygen abundance relative to nitrogen will decrease the N/O ratio in 
these galaxies with higher sSFRs.  Therefore, an anti-correlation between the 
sSFR and N/O ratio is further evidence that oxygen is produced in higher-mass 
stars than those which synthesize nitrogen.  A large-scale environmental 
influence on the relationship between the sSFR and the N/O ratio would manifest 
as a difference in the slopes of the local linear regressions.  With the 
influence of the scatter on these approximations, it is apparent that there is 
no significant influence from the large-scale environment on the relationship 
between N/O and sSFR.

%%%%%%%%%%%%%%%%%%%%%%%%%%%%%%%%%%%%%%%%%%%%%%%%%%%%%%%%%%%%%%%%%%%%%%%%%%%%%%%%
%
%    DISCUSSION
%
%%%%%%%%%%%%%%%%%%%%%%%%%%%%%%%%%%%%%%%%%%%%%%%%%%%%%%%%%%%%%%%%%%%%%%%%%%%%%%%%
\section{Large-scale environmental influence}

We see small, statistically significant shifts in the nitrogen and neon 
gas-phase abundances and the N/O ratio studied as a function of the large-scale 
environment, implying that the large-scale environment has a minor influence on 
the chemical abundances of star-forming dwarf galaxies.  Previous work by 
\cite{Douglass17b} suggests that the oxygen abundance (O/H), nitrogen abundance 
(N/H), and N/O ratio depend on a galaxy's environment.  Work by \cite{Shields91} 
finds no shift in the N/O ratio between cluster and field galaxies, though they 
do find that cluster galaxies have higher metallicities than field galaxies.  
\cite{Contini02} and \cite{Pilyugin02} find a statistically insignificant shift 
in the N/O ratio between cluster and field galaxies, where cluster galaxies have 
lower N/O ratios than field spiral galaxies.  The shifts seen in each of these 
latter three sources are opposite to what we observe in this paper, though these 
previous studies concentrate on the galaxies in the Virgo cluster, which are 
more massive than our dwarf galaxy ($M_r > -17$) sample.  On average, we find 
that star-forming void dwarf galaxies have $\sim$10\% higher neon abundances, 
$\sim$5\% lower nitrogen abundances, and $\sim$7\% lower N/O ratios than 
star-forming dwarf galaxies in denser regions.

As outlined in \cite{Douglass17a}, there have been numerous previous studies 
that investigate the influence of the environment on the metallicity of a 
galaxy, resulting in mixed conclusions.  When a difference in the metallicity 
was attributed to the environment \citep[as in][for example]{Pustilnik06,
Cooper08,Pustilnik11b,Pustilnik14,SanchezAlmeida16}, it was found that those 
galaxies with lower metallicities preferentially reside in less dense regions.  
This is the opposite of the insignificant shift seen in Fig. \ref{fig:met1sig} 
in the star-forming dwarf galaxies, although our results are not a direct 
comparison with their conclusions.  (Due to our requirement of the [\ion{O}{3}] 
$\lambda$4363 auroral line, we are not able to probe the high-metallicity 
regime.)  We observe that there is a statistically insignificant shift toward 
higher average metallicities in void dwarf galaxies compared to dwarf galaxies 
in denser regions.

\subsection{Higher metallicities in void dwarf galaxies}

We posit that if the shift toward slightly higher metallicities in the 
star-forming void dwarf galaxies was significant, it would be due to a 
large-scale environmental effect on the ratio of a galaxy's dark matter halo 
mass to stellar mass ($M_\text{DM}/M_*$).  \cite{Goldberg04} show that 
gravitational clustering within a void proceeds as if in a very low-density 
universe, where the growth of gravitationally bound dark matter halos ends 
relatively early.  Afterwards, there is relatively little interaction between 
the void galaxies because of the lower density and faster local Hubble 
expansion.  Simulations by \cite{Jung14} and \cite{Tonnesen15} show that, for a 
fixed dark matter halo mass, the stellar masses of central galaxies located in 
voids are smaller than those of central galaxies living in denser regions.  The 
$\Lambda$CDM cosmology predicts that galaxies formed in voids will be retarded 
in their star formation when compared to those in denser environments.  
Therefore, void dwarf galaxies could have higher $M_\text{DM}/M_*$ ratios than 
dwarf galaxies in denser regions.  Indeed, \cite{Tojeiro17} find that the ratio 
of stellar mass to halo mass increases with the large-scale environmental 
density.

If this were the case, then the potential well and virial radius of the void 
galaxies would be large enough to retain more of the heavy elements that are 
blown from the ISM to the CGM of a galaxy (e.g., from a supernova).  The 
simulation results of \cite{Tonnesen15} find that, for central galaxies with 
halo masses between $10^{11}$ and $10^{12.9}$, void galaxies have $\sim$10\% 
larger ratios of dark matter halo mass to stellar mass compared with wall 
galaxies at $z = 0$.  In wall galaxies, more of these heavy elements can escape 
the dwarf galaxy, while in void galaxies they are confined to the CGM and 
eventually fall back onto the galaxy's ISM.  If star-forming void dwarf galaxies 
are able to retain more oxygen relative to their hydrogen abundance, then they 
will reach the critical value of O/H for secondary nitrogen production (via the 
CNO cycle) earlier than the star-forming wall dwarf galaxies, for a given 
stellar mass.  We see evidence of this in Fig. \ref{fig:M_NO}, where the N/O 
plateau for the void dwarf galaxies exists for stellar masses 
$\log(M_*/M_\odot) \lesssim 7.7$.  In contrast, the N/O plateau for the wall 
dwarf galaxies exists for stellar masses $\log(M_*/M_\odot) \lesssim 7.8$.

If void dwarf galaxies have higher $M_\text{DM}/M_*$ ratios than dwarf galaxies 
in denser environments, we would expect to see an environmental influence on 
other characteristics of these galaxies.  First, the ratio of neutral hydrogen 
mass to stellar mass of the dwarf galaxies would be higher in star-forming void 
galaxies than star-forming wall galaxies, under the assumption that neutral 
hydrogen traces dark matter.  We see this effect both in Fig. 9 of 
\cite{Moorman16} and in Fig. \ref{fig:HI_hist}.  Second, the \ion{H}{1} mass 
function should shift less from wall to void galaxies than the shift seen in the 
luminosity function.  Indeed, \cite{Moorman16} finds a shift of characteristic 
\ion{H}{1} mass by a factor of 1.4 in the \ion{H}{1} mass function, while 
\cite{Hoyle05} measures a shift by a factor of 2.5 in luminosity in the 
luminosity function between these two environments.  The fact that we find an 
insignificant shift in the metallicity of dwarf galaxies between the void and 
denser environments indicates that the difference in the $M_\text{DM}/M_*$ 
ratios between the two environments is not that significant (or it does not have 
much affect on the chemical composition of a galaxy's ISM).

\subsection{Lower N/O ratios in void dwarf galaxies}

As we see in Fig. \ref{fig:NOratio}, star-forming void dwarf galaxies have a 
lower N/O ratio than wall dwarf galaxies.  This strengthens the preliminary 
results found in \cite{Douglass17b} and the simulation results of \cite{Cen11}, 
indicating that void galaxies are retarded in their star formation and that 
cosmic downsizing might depend on the large-scale environment.  As suggested by 
\cite{vanZee06a}, a galaxy with a declining SFR at late times (a wall galaxy) 
will have a higher N/O ratio than a galaxy with a constant SFR (a void galaxy); 
the ongoing star formation in the void galaxies will release more oxygen into 
the ISM, decreasing their N/O ratio relative to the galaxies with declining 
SFRs.  This concept is supported by the color--N/O diagram in Figures 
\ref{fig:ur} and \ref{fig:gr}, where the bluer galaxies have lower N/O ratios.  
The correlations between color and the N/O ratio found in \cite{vanZee06a}, 
\cite{Berg12}, \cite{Douglass17b}, and this work are a result of declining SFRs 
\citep{vanZee06a}.  The average lower N/O ratios that we see in star-forming 
void dwarf galaxies may be observational evidence that cosmic downsizing is 
environmentally dependent.

\subsection{Extremely low-metallicity dwarf galaxies}

Previous studies have either hypothesized \citep{Pustilnik06,Pustilnik11b,
Pustilnik13} or concluded \citep{Filho15,SanchezAlmeida16} that low-metallicity 
objects preferentially reside in low-density environments.  \cite{Douglass17a} 
find that, out of 135 dwarf galaxies studied, there is no difference in the 
fraction of low-metallicity dwarf galaxies that reside in voids and denser 
regions.  Of the 1722 dwarf galaxies we analyze, 82 have extremely low 
metallicity values (\OH $\leq 7.6$).  Of these 82 low-metallicity dwarf 
galaxies, 46 are found in voids (approximately 5\% of the star-forming void 
dwarf galaxy population studied) and 36 are located in denser regions (5\% of 
the star-forming wall dwarf galaxy population studied).  These population 
fractions do not support the existence of a special population of extremely 
metal-poor dwarf galaxies in voids.

We find that these 82 extremely metal-poor galaxies have comparably lower 
nitrogen abundances when compared to the total star-forming dwarf galaxy 
population studied.  From Figures \ref{fig:ur} and \ref{fig:gr}, it is apparent 
that these extremely metal-poor star-forming dwarf galaxies cover the entire 
color range of dwarf galaxies in this study.  Fig. \ref{fig:SFR} also shows that 
the sSFRs and sSFRs conform with the observed relationship between the N/O ratio 
and the sSFR.

One possible interpretation of these extremely metal-poor dwarf galaxies is that 
they are green pea galaxies.  A class of luminous compact galaxies, 
\cite{Cardamone09} identified these objects to be star-bursting galaxies at 
higher redshifts.  (Their color is a result of unusually large [\ion{O}{3}] 
$\lambda 5007$ emission line fluxes redshifted into the $r$-band of SDSS.)  
These galaxies are found to have low metallicities for their stellar mass; their 
N/O ratios are high for their estimated metallicities.  They typically have 
sSFRs between $10^{-7}$ and $10^{-9}$ yr$^{-1}$ \citep{Izotov11}.  While some of 
the extremely metal-poor galaxies we identify also have high N/O ratios for 
their metallicities, they have sSFRs between $10^{-10}$ and $10^{-8}$ yr$^{-1}$, 
so only some of them might be green pea galaxies.

The few extremely metal-poor galaxies with high N/O ratios could be \ion{H}{2} 
regions contaminated by shock waves.  As described by \cite{Peimbert91}, shock 
waves will increase the intensities of auroral lines.  Since we estimate the 
temperature of the \ion{H}{2} regions with the ratio of [\ion{O}{3}] 
$\lambda$4363 to [\ion{O}{3}] $\lambda \lambda$4959,5007, this would result in 
an overestimation of the temperature and an underestimate of the oxygen 
abundance.  The nitrogen abundance would also be underestimated, but to a lesser 
extent (as it is less sensitive to the electron temperature); this combination 
results in an overestimate of the N/O ratio.  We would see evidence of shock 
waves dominating the spectra of these galaxies in the BPT diagram; because each 
of these galaxies resides in the star-forming region of the BPT diagram, shock 
waves are an unlikely explanation for those extremely metal-poor dwarf galaxies 
with high N/O ratios.

%%%%%%%%%%%%%%%%%%%%%%%%%%%%%%%%%%%%%%%%%%%%%%%%%%%%%%%%%%%%%%%%%%%%%%%%%%%%%%%%
%
%    CONCLUSION
%
%%%%%%%%%%%%%%%%%%%%%%%%%%%%%%%%%%%%%%%%%%%%%%%%%%%%%%%%%%%%%%%%%%%%%%%%%%%%%%%%
\section{Conclusions}

We estimate the gas-phase oxygen, nitrogen, and neon abundances and the N/O and 
Ne/O ratios of star-forming dwarf galaxies in SDSS DR7 using the direct $T_e$ 
method and spectroscopic line flux measurements as reprocessed in the MPA-JHU 
catalog.  We expand upon the previous work of \cite{Douglass17a, Douglass17b} by 
deriving a relation between the doubly ionized oxygen and total oxygen abundance 
in star-forming dwarf galaxies; removing the dependence on the [\ion{O}{3}] 
$\lambda$3727 doublet, this relation allows us to probe those dwarf galaxies at 
$z < 0.02$ in SDSS DR7.  The 1722 dwarf galaxies analyzed indicate that the 
large-scale environment only slightly influences their chemical evolution: 
star-forming void dwarf galaxies have 10\% higher neon abundances, lower 
nitrogen abundances by an average of 5\%, and 7\% lower N/O ratios compared to 
star-forming dwarf galaxies in denser regions.  The large-scale 
($\sim$10 $h^{-1}$Mpc) environment has a small impact on the chemical evolution 
of star-forming dwarf galaxies.

In addition, we also look at the relationship between the metallicity and the 
N/O and Ne/O ratios and other physical characteristics of our star-forming dwarf 
galaxy sample.  There is very little relationship between the Ne/O ratio and the 
metallicity of the star-forming dwarf galaxies, matching expectations since both 
neon and oxygen are produced in $\alpha$ processes.  In the relationship between 
N/O and O/H, all our dwarf galaxies reside on the so-called ``nitrogen 
plateau,'' where the N/O ratio is predicted to be independent of the gas-phase 
oxygen abundance for metallicities \OH $< 8.5$.  We also find a plateau in our 
relationship between the stellar mass and N/O ratio.  Most of our star-forming 
dwarf galaxies follow the typical mass-metallicity relation.  There is no 
relationship between the metallicity and \ion{H}{1} mass of the galaxies, but 
the N/O ratio decreases with increasing \ion{H}{1} mass for fixed stellar mass.  
The star-forming dwarf galaxies exhibit an increase in the metallicity (O/H) and 
N/O ratio with increasing color (both $u-r$ and $g-r$).  We see very little 
correlation with SFR for either metallicity (O/H) or N/O ratio of dwarf 
galaxies, but the metallicity and N/O ratio decrease with increasing sSFR.  
Beyond the slight large-scale environmental influence on the chemical abundance 
distributions in the sample of star-forming dwarf galaxies, we do not observe 
any significant differences in the star-forming void and wall dwarf galaxies of 
any of these relationships.

We surmise that the difference in the distribution of the N/O ratio seen in the 
sample of star-forming dwarf galaxies is due to a slight large-scale 
environmental influence on their formation history and evolution.  The slight 
shifts in the gas-phase oxygen and neon abundance distributions could be 
observational evidence for delayed star formation in void galaxies when compared 
to those in denser regions.  This would result in a smaller ratio of stellar 
mass to dark matter halo mass in void galaxies than in wall dwarf galaxies, as 
predicted in simulations by \cite{Jung14} and \cite{Tonnesen15}.  Simulations 
looking at how the retention fraction of supernovae ejecta depends on the halo 
mass or dark matter potential would be useful in understanding how significant 
the ratio of dark matter to stellar mass is on the gas-phase chemical abundances 
of a galaxy.  If the void galaxies are retaining more oxygen as a result of 
their deeper potential wells, then they will be able to commence the synthesis 
of secondary nitrogen earlier, as is seen in the mass--N/O relation in Fig. 
\ref{fig:M_NO}.  In addition, the shift toward lower N/O ratios in the 
star-forming void dwarf galaxies may be evidence that cosmological downsizing is 
environmentally dependent.  Our results provide evidence for delayed, ongoing 
star formation in void dwarf galaxies whose dark matter halos ceased coalescing 
earlier than for dwarf galaxies in denser regions.

No special population of extremely metal-poor star-forming dwarf galaxies is 
found in the voids, as we note an equal fraction of low-metallicity dwarf 
galaxies in both the voids and denser regions.

%%%%%%%%%%%%%%%%%%%%%%%%%%%%%%%%%%%%%%%%%%%%%%%%%%%%%%%%%%%%%%%%%%%%%%%%%%%%%%%%
%
%    ACKNOWLEDGEMENTS
%
%%%%%%%%%%%%%%%%%%%%%%%%%%%%%%%%%%%%%%%%%%%%%%%%%%%%%%%%%%%%%%%%%%%%%%%%%%%%%%%%
\acknowledgements
The authors would like to thank Lisa Kewley for her support in estimating the 
ionization parameter, and we thank Stephen W O'Neill Jr. for his help fitting 
piecewise linear functions.  We also extend our appreciation to the anonymous 
referee for their insightful comments and suggestions.  K.A.D. and M.S.V. 
acknowledge support from NSF grant AST-1410525.  R.C. acknowledges support from 
NSF grant AST-1515389.

Funding for the SDSS and SDSS-II has been provided by the Alfred P. Sloan 
Foundation, the Participating Institutions, the National Science Foundation, the 
U.S. Department of Energy, the National Aeronautics and Space Administration, 
the Japanese Monbukagakusho, the Max Planck Society, and the Higher Education 
Funding Council for England.  The SDSS website is \emph{http://www.sdss.org/}.

The SDSS is managed by the Astrophysical Research Consortium for the 
participating institutions.  The participating institutions are the American 
Museum of Natural History, Astrophysical Institute Potsdam, University of Basil, 
University of Cambridge, Case Western Reserve University, University of Chicago, 
Drexel University, Fermilab, the Institute for Advanced Study, the Japan 
Participation Group, Johns Hopkins University, the Joint Institute for Nuclear 
Astrophysics, the Kavli Institute for Particle Astrophysics and Cosmology, the 
Korean Scientist Group, the Chinese Academy of Sciences (LAMOST), Los Alamos 
National Laboratory, the Max Planck Institute for Astronomy (MPIA), the 
Max Planck Institute for Astrophysics (MPA), New Mexico State University, Ohio 
State University, University of Pittsburgh, University of Portsmouth, Princeton 
University, the United States Naval Observatory, and the University of 
Washington.

%%%%%%%%%%%%%%%%%%%%%%%%%%%%%%%%%%%%%%%%%%%%%%%%%%%%%%%%%%%%%%%%%%%%%%%%%%%%%%%%
%    BIBLIOGRAPHY
%%%%%%%%%%%%%%%%%%%%%%%%%%%%%%%%%%%%%%%%%%%%%%%%%%%%%%%%%%%%%%%%%%%%%%%%%%%%%%%%

\bibliographystyle{aasjournal}
\bibliography{Doug0912_sources}

\end{document}